\documentclass[graybox]{svmult}
\usepackage{mathptmx}       % selects Times Roman as basic font
\usepackage{helvet}         % selects Helvetica as sans-serif font
\usepackage{courier}        % selects Courier as typewriter font
\usepackage{type1cm}        % activate if the above 3 fonts are
\usepackage{makeidx}         % allows index generation
\usepackage{graphicx}        % standard LaTeX graphics tool
\usepackage{multicol}        % used for the two-column index
\usepackage[bottom]{footmisc}% places footnotes at page bottom

\usepackage{color}
\usepackage{hyperref}
\definecolor{keywordcolor}{rgb}{0,0,0}

\usepackage{siunitx}

\definecolor{mygreen}{rgb}{0,0.5,0} 
\definecolor{myorange}{rgb}{1.,0.5,0} 
\definecolor{myblue}{rgb}{0,0,0} 
\definecolor{mymagenta}{cmyk}{0,1,0,0.12}

\newcommand{\btext}[1]{{\color{myblue}#1}}

\newcommand{\ntext}[1]{{\color{black}#1}}

\newcommand{\relphase}{\phi_{\rm rel}}

\newcommand{\supiso}{^{({\rm iso})}}
\newcommand{\bI}{{\bf I}}

\newcommand{\bJ}{{\bf J}}
\newcommand{\bB}{{\bf b}}
\newcommand{\ket}[1]{|#1\rangle}

\newcommand{\rhoin}{\rho}
\newcommand{\omNOON}{{\omega_{\rm \noon}}}

\newcommand{\aout}{a_{\rm out}}

\newcommand{\noon}{{NooN}~}

\usepackage{bbold}
\usepackage{amsmath}
\usepackage{amssymb}

\makeindex             % used for the subject index
\begin{document}

\title*{Generation, characterization and use of  atom-resonant indistinguishable photon pairs}
%\title*{Spectrally pure, atom-resonant indistinguishable photon pairs using atomic  filters}
\author{Morgan W. Mitchell}
\institute{ICFO-Institut de Ciencies Fotoniques, Av. Carl Friedrich Gauss, 3, 08860 Castelldefels, Barcelona, Spain \\
ICREA-Instituci\'{o} Catalana de Recerca i Estudis Avan\c{c}ats, 08015 Barcelona, Spain \at  \email{morgan.mitchell@icfo.es}}

\maketitle

%\abstract*{Each chapter should be preceded by an abstract (10--15 lines long) that summarizes the content. The abstract will appear \textit{online} at \url{www.SpringerLink.com} and be available with unrestricted access. This allows unregistered users to read the abstract as a teaser for the complete chapter. As a general rule the abstracts will not appear in the printed version of your book unless it is the style of your particular book or that of the series to which your book belongs.
%Please use the 'starred' version of the new Springer \texttt{abstract} command for typesetting the text of the online abstracts (cf. source file of this chapter template \texttt{abstract}) and include them with the source files of your manuscript. Use the plain \texttt{abstract} command if the abstract is also to appear in the printed version of the book.}

\abstract{
%\newline\indent
%\newline\indent
We describe the generation of atom-resonant indistinguishable photon pairs using nonlinear optical techniques, their spectral purification using atomic filters, characterization using multi-photon interference, and application to quantum-enhanced sensing with atoms.  Using either type-I or type-II cavity-enhanced spontaneous parametric down-conversion, we generate pairs of photons in the resonant modes of optical cavities with linewidths comparable to the 6 MHz natural linewidth of the D$_1$ line of atomic rubidium.  The cavities and pump lasers are tuned so that emission occurs in a mode or a pair of orthogonally-polarized modes that are resonant to the D$_1$ line, at 794.7 nm.  The emission from these frequency-degenerate modes is separated from other cavity emission using ultra-narrow atomic frequency filters, either a Faraday anomalous dispersion optical filter (FADOF) with a 445~MHz linewidth and 57 dB of out-of-band rejection or an induced dichroism filter with an 80 MHz linewidth and $\ge$~35~dB out-of-band rejection.  Using the type-I source, we demonstrate interference of photon pair amplitudes against a coherent state and a new method for full characterization of the temporal wave-function of narrow-band photon pairs.  With the type-II source we demonstrate high-visibility super-resolving interference, a high-fidelity atom-tuned \noon state, and quantum enhanced sensing of atoms using indistinguishable photon pairs. 
%for frequency-degenerate photons, resonating the production of frequency-degenerate photon pairs.  
%
%of pairs of atom-resonant photons, i.e. with mean frequency chosen near to the D$_1$ transition of atomic rubidium.  The photons of the pair are  identical in frequency, spatial mode, arrival time distribution, and in one of the two setups, indistinguishable in polarization as well.  These photon pairs are both capable of multi-photon interference, for example the Hong-Ou-Mandel effect or interference of photon pair amplitudes from distinct sources.  We use cavity-enhanced spontaneous parametric down-conversion, using a cavities with bandwidths similar to the 6 MHz natural linewidth of the atomic transition, and ultra-narrow atomic filters, either a Faraday anomalous dispersion optical filter (FADOF) or an induced dichroism filter to remove the output from all but one mode of the cavity.  Using these photon pairs we demonstrate interference of photon pair amplitudes relative to a coherent state, full characterization of the temporal wave-function of narrow-band photon pairs, and quantum enhanced sensing of atoms using atom-tuned ``NooN'' states. 
%%Please use the 'starred' version of the new Springer \texttt{abstract} command for typesetting the text of the online abstracts (cf. source file of this chapter template \texttt{abstract}) and include them with the source files of your manuscript. Use the plain \texttt{abstract} command if the abstract is also to appear in the printed version of the book.
}

\section{Introduction}
\label{sec:Introduction}

Interference of indistinguishable photons is one of the most striking non-classical phenomena.  When two indistinguishable photons meet at a beamsplitter, each entering from a different port, an interference of two-photon amplitudes leads them to ``coalesce'' and to exit in the same direction, even though there is no force causing them to interact \cite{HongPRL1987,ShihPRL1988}.  This and similar multi-photon interference effects have been harnessed for quantum teleportation \cite{BouwmeesterN1997}, entanglement swapping \cite{PanPRL1998}, linear optics quantum information processing \cite{OBrienN2003}, quantum-enhanced sensing \cite{MitchellN2004}, and quantum simulation \cite{AaronsonARX2010}, among other applications.  A similar effect concerns the interference of indistinguishable photon pairs sharing a single mode against other sources of photon pairs.  This is the  mechanism by which squeezed vacuum, which consists of pairs of indistinguishable photons, alters the noise distribution in an interferometer \cite{WuPRL1986}, an phenomenon central to quantum-enhanced sensing \cite{LIGONP2011, AasiNP2013} and continuous-variable quantum information processing \cite{BraunsteinBOOK2010}.

Most of \btext{the above-mentioned} applications were developed with broadband photons that, due to a strong spectral mismatch, could at best interact inefficiently with atoms or other material systems.  Generation of indistinguishable, atom-resonant photons is an attractive goal if we wish to interact states exhibiting non-classical interference with atomic quantum information processors (see chapters by Leuchs \& Sondermann, Piro \& Eschner, and \btext{Slodi\v{c}ka, H\'etet, Hennrich \& Blatt}), atomic quantum memories (see chapter by Chuu \&  Du), or atomic sensors \cite{WolfgrammPRL2010, KoschorreckPRL2010a, KoschorreckPRL2010b, SewellPRL2012, SewellPRX2014}.  Using cavity-QED methods, atom-resonant indistinguishable photons can be produced with exquisite control over their wave-functions (see the chapter by A. Kuhn), but at a high cost in system complexity.  

In this chapter we describe generation of pairs of indistinguishable photons using cavity-enhanced spontaneous parametric down-conversion (CESPDC) \cite{PredojevicPRA2008, WolfgrammOE2008, WolfgrammJOSAB2010} and extremely narrow-band optical filters \cite{ZielinskaOE2014,ZielinskaOL2012} to select the atom-resonant component of the emission.  
We describe also applications: using pairs of indistinguishable high-coherence photons, we demonstrate a full measurement of the two-photon temporal wave function, including both amplitude and phase \cite{BeduiniPRL2014}, generate an atom-tuned \noon state \cite{WolfgrammJOSAB2010}, and use this state to probe an atomic magnetometer, demonstrating the use of indistinguishable photon pairs for sensing of an atomic system beyond the standard quantum limit \cite{WolfgrammNPhot2013}.  

\subsection{CESPDC Sources}

Spontaneous parametric down-conversion (SPDC) is a proven method to generate non-classical states of light, including single-mode and two-mode squeezed states, entangled photon pairs, and states with multiple pairs of photons.  Although photon production by SPDC is probabilistic, heralded single photons, heralded entangled states, and heralded versions of more exotic states including so-called ``Schr\"odinger kitten'' states, can be generated from SPDC output using single-photon detectors to indicate when the source has produced the desired state.  Quantum states can also be generated ``on demand'' if active elements are also incorporated (see the chapter by Yoshikawa \& Furusawa).  

A strength of SPDC is the simplicity of the parametric interaction, which can produce very pure correlations in frequency, polarization, and photon number. In contrast, parametric interactions do not naturally produce narrowband output, and some engineering is required to generate SPDC output that is bandwidth-compatible with atomic resonance lines or other material spectroscopic features.  Due to the strong frequency correlations, narrow-band filtering of one SPDC output can be used to select desired frequency content in the other output, as in the chapter by Piro \& Eschner.  Placing the SPDC process inside a resonant cavity, ``cavity-enhanced SPDC,'' (CESPDC) \cite{OuPRL1999} gives a resonant enhancement to photon pairs coinciding with the cavity modes while suppressing those frequencies that do not fit.  Cavity-enhancement techniques often benefit from techniques such as group-velocity mismatching and even backward-wave phase-matching that further restrict the SPDC output (see the chapters by  Chuu \& Du, de Riedmatten \&  Afzelius and Zhao, Bao, Zhao \& Pan).

In this chapter we will describe two very similar CESPDC sources, the first using type-I phase matching, and thus capable of producing fully-degenerate photon pairs, i.e. produced in the same cavity mode and thus with the same frequency, polarization, and spatial mode.  The second CESPDC source uses type-II phase matching, and is tuned so that one $H$-polarized mode, and one $V$-polarized mode are both resonant at half the frequency of the pump.  In this way, we generate photon pairs that are indistinguishable in all respects except for polarization.  We will sometimes refer to these devices as sub-threshold optical parametric oscillators (OPOs), the term usually used in frequency conversion and continuous-variable quantum optics for these same devices.  % known as a Faraday Anomalous-Dispersion Optical Filter (FADOF).  

%This places stringent requirements on the performancThis means that the FADOF - with its 445~MHz bandwidth - can successfully filter all the nondegenerate modes, leaving only the photons in the degenerate mode, which are then fully indistinguishable, as they share the same spatial mode, frequency and polarization.

\subsection{Atomic frequency filters}
\label{sec:FADOF}

%Here we describe the use of atomic filters to select photons of a specific, atom-resonant frequency from CESPDC.  
Atomic filters use atomic media, usually atomic vapors, to generate intrinsically narrow-band spectral features, and typically achieve transmission bandwidths from a few MHz to a few GHz.  The ``traditional'' uses of atomic filters are in astronomy \cite{OhmanSOA1956}, \btext{laser ranging and surveying}, and daylight optical communications, where they are used to detect signals at specific frequencies while giving excellent blocking, better than 1:$10^5$, over a very wide rejection band.  As we will describe, these features combine well with the output of our CESPDC source, which has a linewidth of only a few MHz, \btext{free-spectral range (FSR)} of a few hundred MHz and emission extending over hundreds of GHz.  Proper matching of an atomic filter to a CESPDC source can then select a single output line while efficiently rejecting the rest of the SPDC output.  

While atomic media naturally present strong and narrow absorption features, what we require   is a filter with narrowband {\em transmission}.  This requires some kind of optical trickery, to convert the absorption resonances into transmission features.  Equally importantly, the filter must have a strong rejection of the unwanted frequencies.  
%Consider that a simple CESPDC source will emit atom-resonant photon pairs from a single cavity longitudinal mode, or pair of modes, while also emitting non-resonant photon pairs on hundreds of other longitudinal modes spread over hundreds of GHz of bandwidth.  To select from this broadband output atom-resonant photon pairs with high spectral purity requires rejecting these other modes with high efficiency.  
While traditional optics, in the form of cascaded optical cavities, can in principle perform this task perfectly, i.e. with arbitrarily high rejection and unit transmission, this approach requires careful mode-matching and active tuning of the filters, and is ultimately limited by the quality of the cavity optics \cite{Neergaard-NielsenOE2007,HaaseOL2009,PalittapongarnpimRSI2012}.  Atomic filters are an interesting option for their stability, multi-mode capability, high transmission, and very high out-of-band rejection, limited by the quality of the polarizers.  To date, atomic filters continue to outperform cavity filters in these figures of merit.

\begin{figure}[t]
 \centerline{\includegraphics[width=0.9\textwidth]{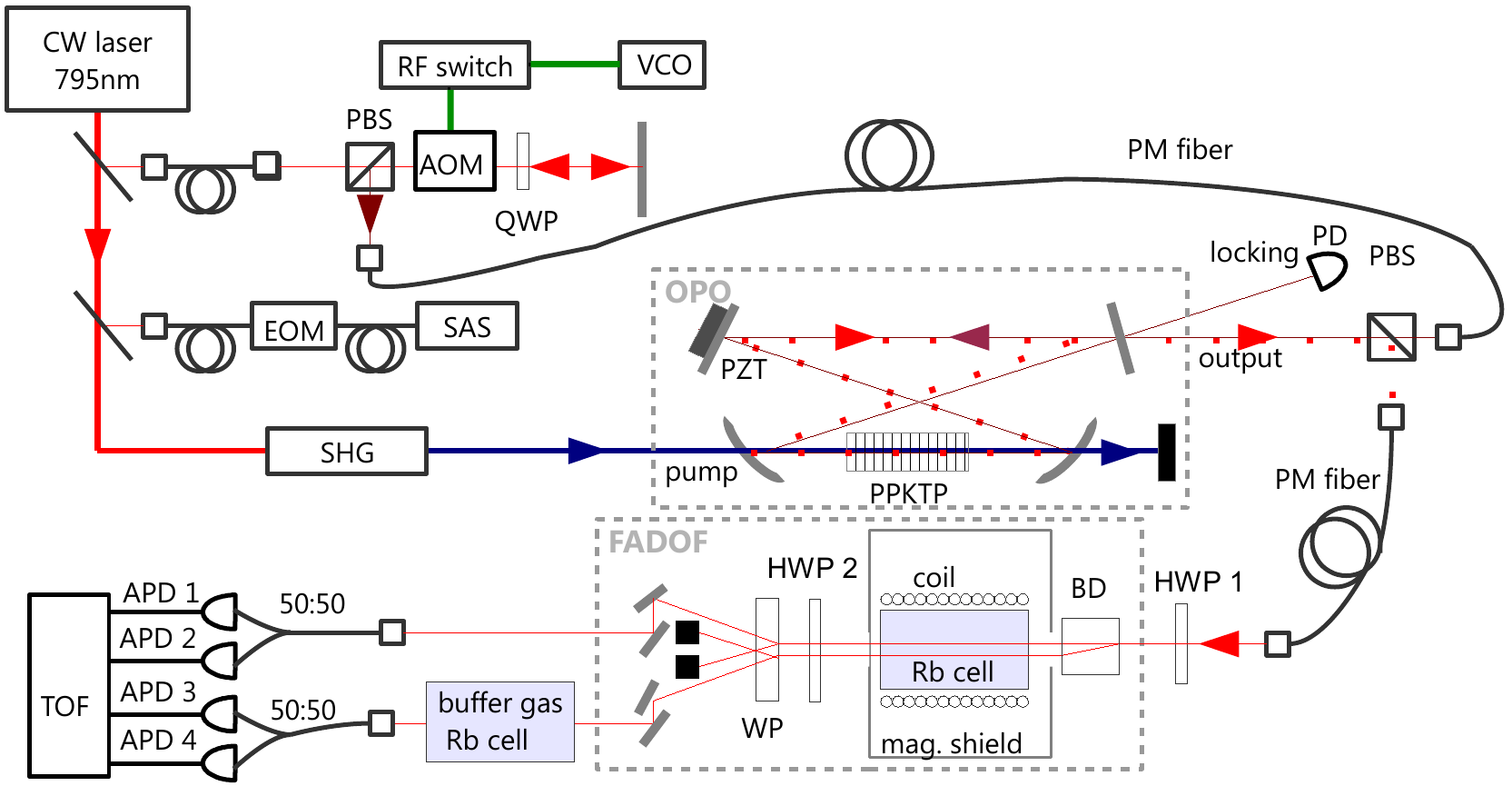}}
\caption{\label{fig:opo}Experimental setup of the type-I OPO, the FADOF filter and detection system. Symbols: PBS: polarizing beam splitter, AOM: acousto-optic modulator, EOM: electro-optic modulator, APD: avalanche photodiode, BD: calcite beam displacer, WP: Wollaston prism, TOF: time-of-flight analyzer, SAS: saturated absorption spectroscopy, VCO: voltage controlled oscillator, PM: polarization maintaining fiber, HWP: half-wave plate, QWP: quarter-wave plate, PD: photodiode}
\end{figure}

\section{Atom-resonant indistinguishable photon pairs in a single mode}

\btext{In this section we describe a series of experiments to generate indistinguishable photon pairs in a single polarization mode, using type-I parametric downconversion.  Subsection \ref{sec:TypeI} describes the source, subsection \ref{sec:RbD1FADOF} describes the atomic filter,  subsection \ref{sec:TypeIPurification} describes measurements of the spectral purity achieved, subsection \ref{sec:BiphotonInterference} demonstrates interference of two-photon amplitudes between a narrow-band SPDC source and a CW laser, and subsection  \ref{sec:Reconstruction} applies this interference to biphoton wave-function measurement.}   Additional details can be found in \cite{ZielinskaOL2012, ZielinskaOE2014,BeduiniPRL2014,ZielinskaARX2014b}.

%Our CESPDC source is a sub-threshold OPO consisting of an optical resonator with a blue-light-pumped $\chi^{(2)}$ nonlinear medium inside.  The OPO generates degenerate narrowband photo pairs near atom-resonant squeezed vacuum in the degenerate cavity mode, i.e., the longitudinal mode with half the pump frequency, but also a far larger number of non-degenerate photon pairs in other modes.  

\subsection{Type-I CESPDC source}
\label{sec:TypeI}
We use a doubly-resonant degenerate OPO \cite{Anacavity} containing a type-I PPKTP crystal, phase-matched for second-harmonic generation from 794.7 nm to 397.4 nm.  A schematic is shown in Fig. \ref{fig:opo}.  A continuous wave external cavity diode laser at 794.7 nm is stabilised at a frequency $\omega_0$, 2.7 GHz to the red of the Rb D$_1$ line centre.  As described in \cite{ZielinskaOL2012}, an electro-optic modulator (EOM) adds sidebands to a saturated spectroscopy absorption signal in order to get an error signal at the desired frequency. We double the laser frequency by cavity-enhanced second harmonic generation using a LBO crystal to generate a 397.4 nm pump beam for the OPO.  The cavity length is stabilized to maintain a TEM00, $V$-polarized cavity mode at frequency $\omega_0$, thus resonating the SPDC production of indistinguishable photon pairs at this frequency. 

With this configuration, photon pairs are generated at all the resonance frequencies of the OPO cavity that fall inside the  150~GHz-wide phase matching envelope of the PPKTP crystal. The OPO output is thus composed of hundreds of frequency modes, each of 8.4~MHz bandwidth, separated by the 501~MHz FSR.  We note that in this type-I scenario, group velocity mismatching techniques are not applicable, because the pairs of generated photons differ only in longitudinal mode.  Filtering can in principle separate the desired, degenerate-mode photon pairs from the background pairs, but the requirements are quite stringent: The filter must be able to distinguish between one cavity mode and the next, separated by the FSR. Moreover, to achieve a high spectral purity in the output, the filter needs a very high out-of-band rejection.  Finally, high transmission efficiency is always desirable so as to not lose the photon pairs.  As we shall see, these requirements are well-matched to a specific kind of atomic frequency filter.

\newcommand{\mycite}{\cite}

\subsection{A FADOF  at the Rb D$_1$ line}
\hyphenation{FADOF FADOFs}
\label{sec:RbD1FADOF}
{
%Ultra-narrow bandwidth optical filters based on resonant atomic susceptibilities 
%are key elements in laser remote sensing (LIDAR), observational astronomy, and 
%free-space communications.  Relative to conventional interference filters, 
%atomic filters offer high background-rejection, mechanical robustness,
%imaging capability, and high transmission.  These same features make them interesting
%for separating the atom-resonant photons from a CESPDC source from the more numerous non-resonant pairs from the same source.  

We now describe a Faraday Anomalous-Dispersion Optical Filter (FADOF) at the D$_1$ line of Rb at 795 nm.  This line, efficiently detected with Si detectors, accessible with a 
variety of laser technologies,
and showing large hyperfine splittings, is a favorite for coherent and quantum
optics with warm atomic vapors.  Applications include electromagnetically-induced
transparency \mycite{EIT4WM}, stopped light \mycite{stopped}, optical
magnetometry \mycite{EITmag,WolfgrammPRL2010},  laser
oscillators \mycite{RbLaser}, polarization squeezing
\mycite{PolSqueezedLvovsky,AghaOE2010}, quantum
memory \mycite{memoryHosseini}, and high-coherence heralded single photons
\mycite{CereOL2009,WolfgrammPRL2011}.

%Here we show that the Rb D$_1$ line provides superior FADOF performance.  
%We demonstrate a FADOF surpassing other atoms and other Rb transitions for key figures of merit,
%including peak transmission $T_{\rm max}$, transmission bandwidth and  equivalent noise bandwidth
%${\rm ENBW}={T^{-1}_{\rm max}}\int T(\nu)d\nu$, where $T(\nu)$ is the filter transmission versus frequency $\nu$ \mycite{RbD2}.  For narrowband signals in broadband noise, a filter achieves the signal-to-noise of an ideal filter with bandwidth ENBW.}

\begin{figure}[htb]
\centerline{
 \includegraphics[width=0.85 \columnwidth]{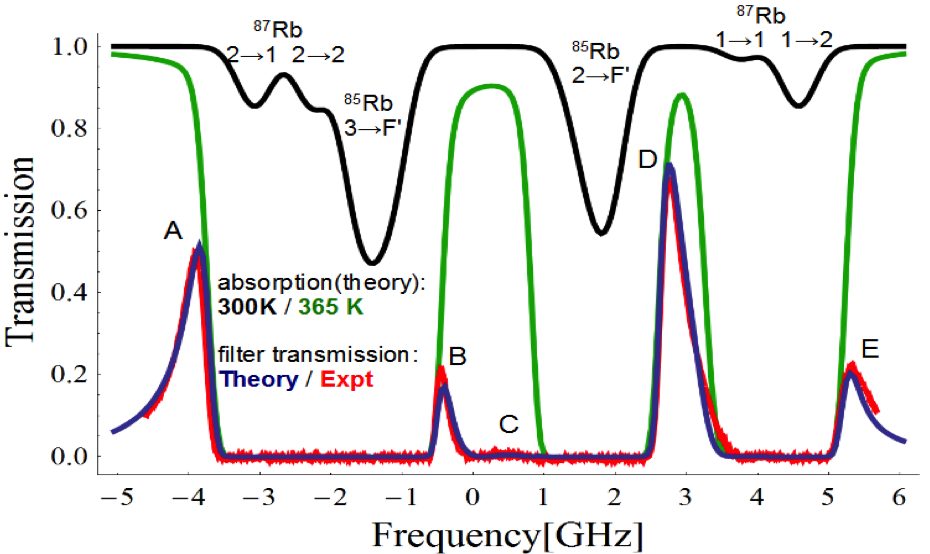}
 }
\caption{\label{fig:Theoretical-absorption-spectrum}
Features of the Rb D$_{1}$ FADOF spectrum for a cell  of 10 cm internal length containing natural abundance rubidium with no buffer gas.  The calculated absorption spectrum at \SI{300}{K} is shown in black, with distinct peaks due to the $^{85}$Rb and $^{87}$Rb transitions.  The calculated spectrum at \SI{365}{K} is shown in green, and shows very strong absorption in three regions.  When this medium is placed between crossed polarizers in the presence of a magnetic field of $B=\SI{4.5}{mT}$, strong Faraday rotation at the line-edges leads to sharp transmission peaks an points A-E.  Calculated transmission spectra are shown in blue, experimental spectra are shown in red.  The feature D, which occurs between two lines, is particularly advantageous, as it is both stronger (higher transmission), and narrower than other lines, because its tail is absorbed by the   $^{87}$Rb $F=1 \rightarrow F'$ transitions.   }
\end{figure}

A FADOF
%, shown schematically in Fig. \ref{fig:Theoretical-absorption-spectrum}, 
consists of an atomic vapor cell between two crossed polarizers, while a
homogeneous magnetic field along the propagation direction induces
circular birefringence in the vapor. The crossed polarizers
block transmission away from the absorption line, while the absorption itself
blocks resonant light.  Between these, Faraday rotation just outside the Doppler
profile can give high transmission for a narrow range of frequencies.
FADOF is simple and robust, but performance depends critically on optical
properties of the atomic vapor.  Fortunately, first-principles
modeling of the atomic vapor agrees very well with experiment, as shown in 
Fig. \ref{fig:Theoretical-absorption-spectrum}.  
Public-domain codes for calculating Rb \cite{ZielinskaOL2012} and Na  \cite{GerhardtNaFadofCode,KieferSR2014}  FADOF spectra are available.

Similar FADOFs have been developed for
several other alkali atom resonances
 -- Cs D$_2$ \mycite{MendersOL1991} and $6S_{1/2}\rightarrow7P_{3/2}$  \mycite{YinPTL1992,WangOE2012} lines, Rb D$_2$ line
\mycite{RbD2report,RbD2}, Rb $5S_{1/2}\rightarrow 6P_{3/2}$ \mycite{LingOL2014}, K  (three lines)  \mycite{KD1},
Na D lines \mycite{Sodium}, and for Ca \mycite{Chan1993}.  Several of these show transmission above 90\%, and/or linewidths below 1 GHz.   Filter figures of merit are shown in Table \ref{tab:comparison}.  An important figure of merit used in the FADOF literature is the Equivalent Noise Bandwidth 
${\rm ENBW}=[\max_\nu T(\nu)]^{-1}\int T(\nu)d\nu$, where $T(\nu)$ is the filter transmission versus frequency $\nu$ \mycite{RbD2}.  For narrowband signals in broadband noise, a filter achieves the signal-to-noise of an ideal filter with bandwidth ENBW.  

{
\begin{table}
\centering
\caption{\label{tab:comparison}
Comparison of reported FADOF transmission for different atoms and wavelengths $\lambda$.  T$_{\rm max}$: peak transmission.   B$_{\rm T}$: full-width at half-maximum bandwidth of main transmission peak. B$_{\rm N}$: Equivalent-noise bandwidth (ENBW).  - value not reported.  Reference   \cite{KD1} shows a K \SI{770}{nm} FADOF curve very similar to K \SI{766}{nm}.    }

\begin{tabular}{cccccccc}
 &  &  & \tabularnewline
\hline
Atom& $\lambda$[nm]  & Ref.  &&& T$_{\rm max}$  &B$_{\rm T}$[GHz]&B$_{\rm N}$[GHz]\tabularnewline
\hline
K & 405 &\mycite{KD1}  &&& 0.93  & 1.2 & 6\tabularnewline
$^{87}$Rb & 420 &\mycite{LingOL2014}  &&& 0.98   & 2.5  & 5.9 \tabularnewline
Ca &423 & \mycite{Chan1993}  &&& 0.55  & 1.5 & - \tabularnewline
Cs &455 &\mycite{YinPTL1992}    &&& 0.96  & 0.9 & 3.3 \tabularnewline
Cs &455 &\mycite{WangOE2012}   &&& 0.86  & 1.5 & - \tabularnewline
Na &589 &\mycite{Sodium}  &&& 0.85  & 1.9& 5.1  \tabularnewline
Na &590 &\mycite{Sodium}   &&& 0.37  & 10.5 & 8.3 \tabularnewline
K &766 &\mycite{KD1} &&& 0.96  & 0.9 & 5   \tabularnewline
K &770 &\mycite{KD1} &&& -  & - & -   \tabularnewline
%K &766 &\mycite{KD2} & -  & - & -   \tabularnewline
Rb &780 & \mycite{RbD2}   &&& 0.93  & 1.3& 4.7 \tabularnewline
Cs &852 &\mycite{MendersOL1991}   &&& 0.90  & 0.6 & - \tabularnewline
\hline
& & & B[mT]  &T[K]  &  &   \tabularnewline
\hline
Rb & 795 & \cite{ZielinskaOL2012} & 18.0   &353 & 0.92 & 0.48 & 2.1\tabularnewline
Rb & 795 & \cite{ZielinskaOL2012} &  5.9  &378 & 0.91 & 1.10 & 2.7  \tabularnewline
Rb & 795 & \cite{ZielinskaOL2012} &   4.5  & 365 & 0.71 & 0.45& 1.2\tabularnewline
Rb & 795 & \cite{ZielinskaOL2012} &   2.0  &345 & 0.04 & 0.32 & 0.8 \tabularnewline
\end{tabular}

\end{table}
}

\btext{When quantified by the ENBW, the Rb D$_1$ line gives superior performance to other species and lines, due to what} appears to be a fortunate accident of the hyperfine splittings.  For either pure $^{85}$Rb or pure $^{87}$Rb, the FADOF transmission at these field strengths shows four peaks, with the strongest ones at the extremes of the spectrum and with long tails.  The strong $^{87}$Rb peaks are visible as peak A and E of Fig. \ref{fig:Theoretical-absorption-spectrum}.  \btext{The strong $^{85}$Rb peaks include the peak D and one at \SI{-2.5}{GHz}, but in the natural-abundance vapor this latter peak is completely obscured by the $^{87}$Rb absorption.}  The long tail of peak D is blocked by absorption from the $^{87}$Rb $F=1\rightarrow F'=1$ transition, improving the ENBW.

\subsubsection{A dual-channel FADOF }

The filter used here is a small modification of that described in Section \ref{sec:RbD1FADOF} and in \cite{ZielinskaOL2012}. 
We take advantage of the multi-mode, imaging property of the FADOF to filter simultaneously two orthogonal polarizations: instead of the crossed polarizers, we use a beam displacer before the cell, so that the two orthogonal polarizations travel along independent parallel paths in the cell.   After the cell we use a Wollaston prism to separate the near-resonant filtered light from the unrotated one. The optical axes of the two polarizing elements are oriented with precision mounts, and an extinction ratio of $1.8 \times 10^{-6}$ is reached. 

Additionally, the setup has been supplemented with a half-waveplate placed before the Wollaston prism (HWP~2 in Fig. \ref{fig:opo}), which enables us to, in effect, turn on and off the filter. In the ``FADOF on'' condition, the waveplate axis is set parallel to the Wollaston axis (and thus the waveplate has no effect on the filter behaviour), the magnetic field is 4.5 mT and the temperature is 365 K.  In the ``FADOF off'' condition, no magnetic field is applied, the temperature of the cell is also 365 K and  HWP~2 is set to rotate the polarization by 90 degrees, in effect swapping the outputs, so that almost all the light is transmitted through the setup {without being filtered}. 

\begin{figure}[htb]
 \centerline{\includegraphics[width=0.8 \textwidth]{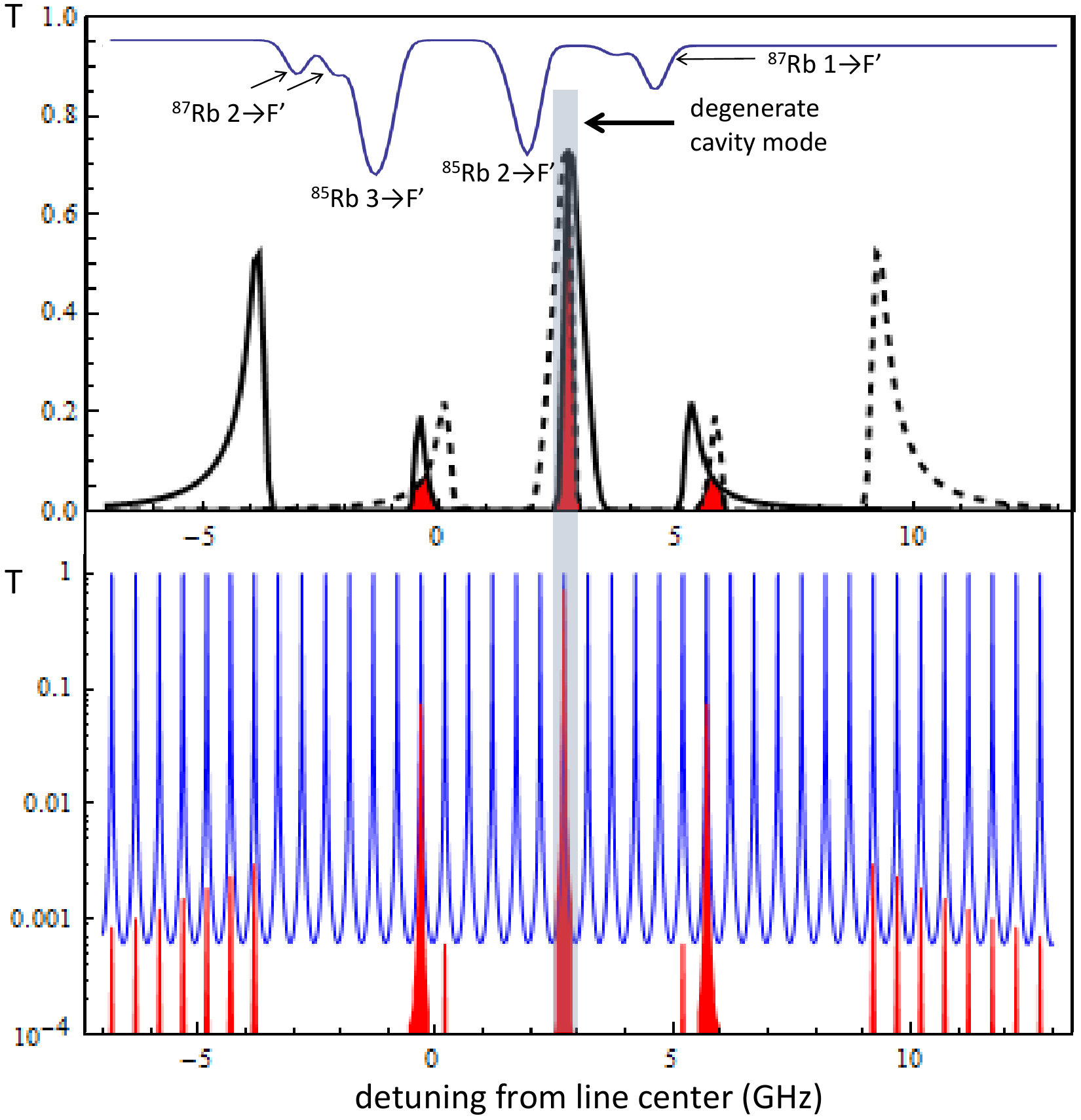}}
\caption{\label{fig:filterandcavity} Spectral matching of FADOF to CESPDC source.  Upper plot: reference transmission spectrum of room temperature natural abundance Rb (blue), filter spectrum (black) and a mirror filter spectrum with respect to the degenerate cavity mode (black dashed). Red shaded regions indicate transmission of correlated photon pairs. Lower plot: cavity output spectrum (blue) and FADOF-filtered cavity spectrum (red). The degenerate cavity mode coincides with the FADOF peak. Both figures have the same frequency scale.}
\end{figure}

We optimized the filter using a common criterion for experiments with photon pairs:  we maximize the ratio of coincidences due to photon pairs belonging to the degenerate mode to coincidences due to other photon pairs. Because of energy conservation, the two photons in any SPDC pair will have frequencies symmetrically placed with respect to the degenerate mode; to prevent the pair from reaching the detectors, it suffices to block at least one of the photons.  In terms of filter performance, this means that it is possible to have near-perfect filtering even with transmission in some spectral windows away from the degenerate mode, provided the transmission is asymmetrical (Fig. \ref{fig:filterandcavity}).  Using this criterion we find the optimal conditions for the filter performance \btext{at the field and temperature values given above.} 
 %in our experiment to be 4.5 mT of magnetic field and the cell temperature of 365 K. 
 The optimum filter performance requires the degenerate mode that should be filtered to coincide with the FADOF transmission peak at a fixed frequency (2.7 GHz to the red from the center of the Rb D$_1$ line).\\

{
}

\subsection{Spectral purification of degenerate photon pairs from type-I CESPDC}
\label{sec:TypeIPurification}

We now show how such an atomic filter can separate indistinguishable, atom-resonant photon pairs from a much stronger broadband background of non-degenerate photon pairs, the natural output of a sub-threshold OPO or CESPDC source.  
%We work with single-mode squeezed vacuum, the starting ingredient for production of ``Schr\"{o}dinger kitten'' states \cite{OurjoumtsevS2006} by photon subtraction, generated
%We work at the D$_1$ line of atomic rubidium at 794.7 nm, a favorite wavelength for atomic quantum memories.  Thulium-doped solid state quantum memories \cite{SaglamyurekN2011, PascualWinterPRB2012} also operate at nearly this same wavelength. 
%Our CESPDC source is a sub-threshold OPO consisting of an optical resonator with a blue-light-pumped $\chi^{(2)}$ nonlinear medium inside.  The OPO generates degenerate narrowband photo pairs near atom-resonant squeezed vacuum in the degenerate cavity mode, i.e., the longitudinal mode with half the pump frequency, but also a far larger number of non-degenerate photon pairs in other modes.  
%We use a modified Faraday rotation anomalous-dispersion optical filter (FADOF) \cite{ZielinskaOL2012} to separate the single-mode squeezed vacuum from these other, co-propagating, modes.  
%Previously, atomic filters have been used to filter single photons \cite{WolfgrammPRL2011} and polarization-distinguishable photon pairs \cite{WolfgrammNPhot2013}, but with lower efficiencies (up to 14\%), incompatible with non-classical continuous-variable states. 

Using the FADOF described in Section \ref{sec:FADOF} we observe 70\% transmission of the degenerate mode through the filter, 
%compatible with 5 dB of squeezing, 
simultaneous with out-of-band rejection by 57~dB, sufficient to reduce the combined non-degenerate emission to a small fraction of the desired, degenerate mode emission.   For comparison, a recently-described monolithic filter cavity achieved 60\% transmission and 45~dB out-of-band rejection \cite{PalittapongarnpimRSI2012}.  
We test the filter by coincidence detection of photon pairs from the OPO output, which provides a stringent test of the suitability for use at the single-photon level.  
We observe for the first time fully-degenerate, near atom-resonant photon pairs, as evidenced by correlation functions and  atomic absorption measurements.  The 96\% spectral purity we observe \btext{surpasses the previous record }
%is the highest yet reported for photon pairs, surpassing the previous record 
of 94\% \cite{WolfgrammPRL2011}, and \btext{is} in agreement with theoretical predictions.  
  
%\section{Experiment}
\label{sec:Experiment}

\subsubsection{Detection}
\label{sec:detection}

The distribution of arrival times of photons in a Hanbury-Brown-Twiss configuration is useful to check that the filter effectively suppresses the non-degenerate modes of the type-I OPO described in the previous section. We collect the OPO output in a polarization maintaining fiber and send it through the filter setup. The filtered light is then coupled into balanced fiber beam splitters that send the photons to avalanche photo-detectors (APDs), connected to a time-of-flight analyzer (TOF) that allows us to measure the {second order correlation function} $G^{(2)}(T)$ (see Fig. \ref{fig:opo}). 

Since we are using single photon detectors, we need to reduce as much as possible the background due to stray light sources in the setup. The main source of background light is the counter-propagating beam that we inject in the OPO in order to lock the cavity length to be resonant at $\omega_0$. We solve this problem using a chopped lock: the experiment switches at 85 Hz  between periods of data acquisition and periods of stabilization.  During {periods of} data acquisition, the AOM is off, and thus no locking beam is present.  During {periods of} stabilization, the AOM is on, and an electronic gate circuit is used to block electronic signals from the APDs, preventing recording of detections due to the locking beam photons. In addition, the polarization of the locking beam is orthogonal to that of the OPO output.

\begin{figure}
\centering
  \includegraphics[width= 0.7\textwidth]{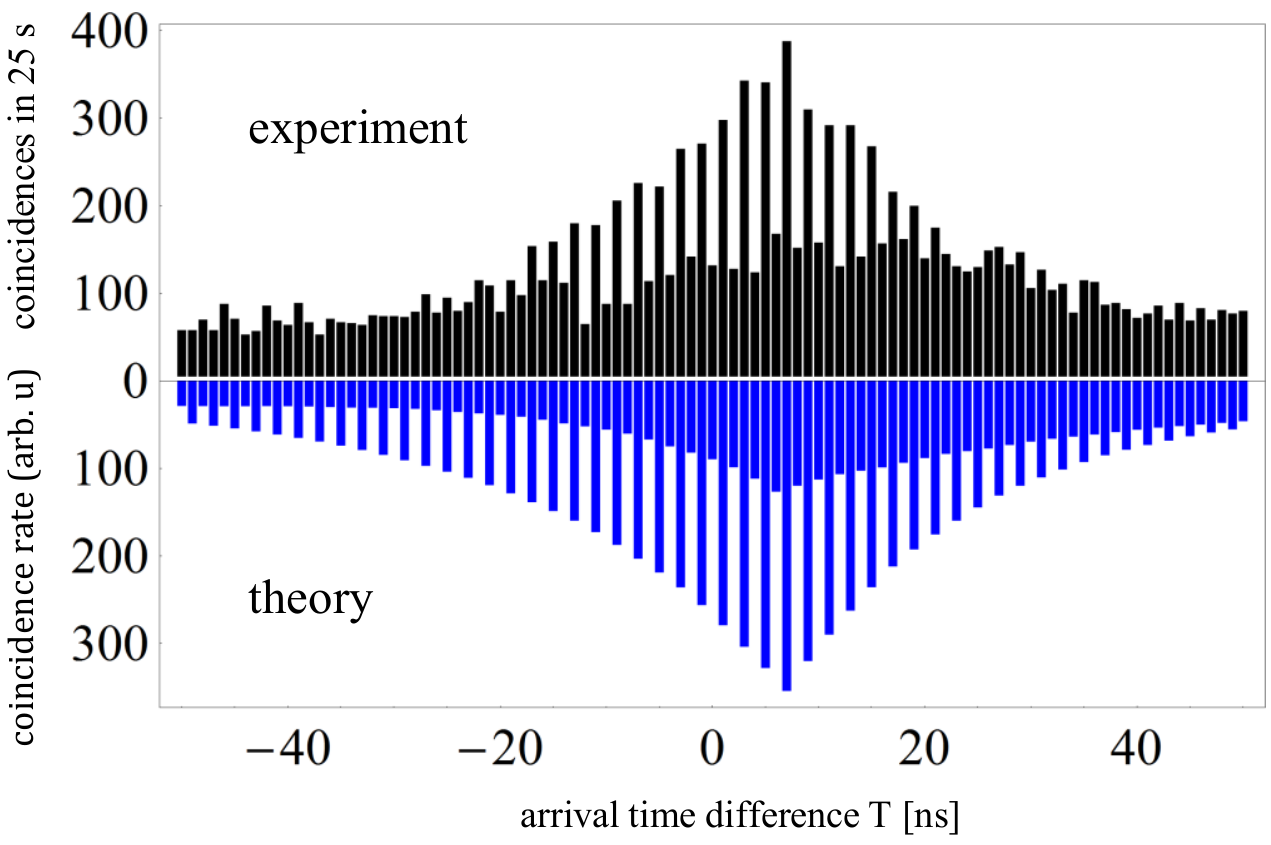}
\caption{\label{fig:fadofoff} Histograms of arrival time differences for FADOF off compared to theoretical model (both include the background due to accidental coincidences and the artefacts resulting from 1 ns resolution of the counting electronics).  
}
\end{figure}

\begin{figure}
\centering
  \includegraphics[width=0.8 \textwidth]{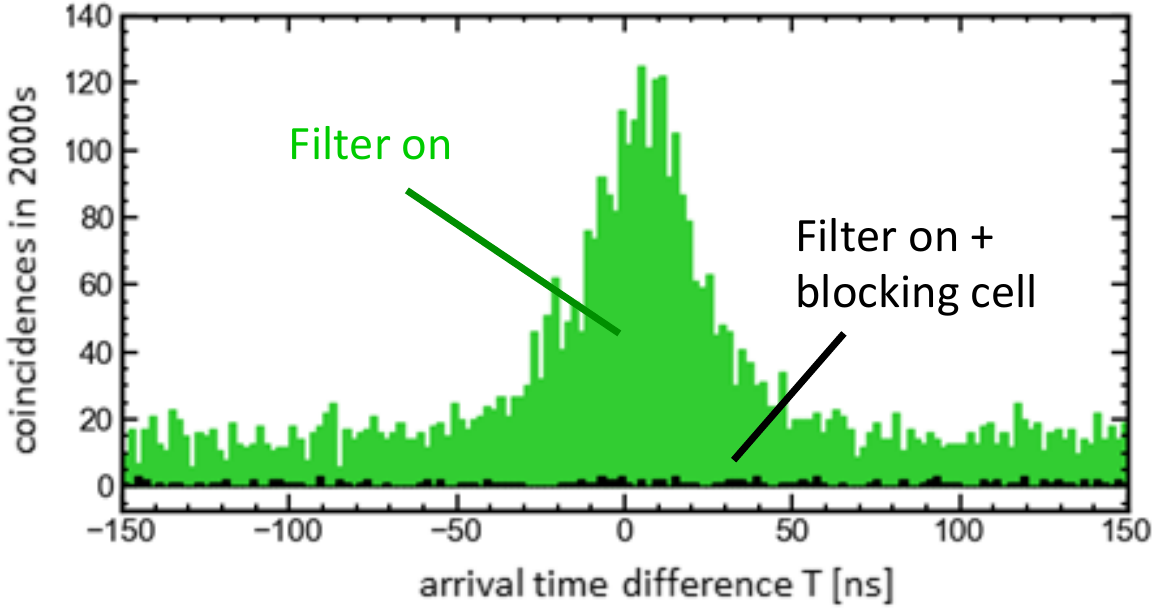}
\caption{Histograms of the differences of arrival times of the photon pairs for FADOF on (green) and FADOF on with hot cell on the path (black bars at bottom of graph, look closely!). No background has been subtracted.
}
\label{fig:fadofon} 
\end{figure}

\subsubsection{Effect of filtering on arrival time distribution}
\label{sec:FilteredArrivals}
The relative arrival-time distribution for photon pairs produced by the CESPDC source is a Dirac comb, with a separation given by the cavity round-trip time, times a double exponential with a time constant given by the cavity ring-down time \cite{KuklewiczPRL2006,ZielinskaARX2014b}.  Here the cavity round-trip time is slightly less than 2 ns, whereas the resolution of our time-to-digital converter is 1 ns.  Evidence for the comb structure is visible in the measured distribution, shown in Fig.  \ref{fig:fadofoff}.  This agrees well with the theoretical expectation, described in the Appendix, as does the 26 ns  full-width at half-maximum (FWHM) of the double-exponential envelope.  When the filter is ``on,'' we expect to see an unmodulated double exponential of the same width.  While the filter blocks the unwanted modes, it has little reshaping effect on the degenerate mode, which is much narrower than, and centred on, the peak of the filter pass-band.  In agreement with these expectations we observe a double-exponential distribution with no visible modulation with the same FWHM width.  This is shown in Fig. \ref{fig:fadofon}.

%Histograms of photon arrival time differences for the ``FADOF off'' and ``FADOF on'' configurations are shown in Figs. \ref{fig:fadofoff} and \ref{fig:fadofon}, respectively.
%We observe that the double exponential full-width at half-maximum (FWHM) corresponds to the predicted one of 26~ns. Moreover, we notice how the comb structure is not present in the ``filter on'' data, as expected if the filter blocks all pairs not in the degenerate mode.  

\subsubsection{Spectral purity}
\label{sec:FADOFSpectralPurity}
\newcommand{\SP}{P_{\rm S}}

According to the theoretical filter spectrum from \cite{ZielinskaOL2012}, we estimate that 98\% of the atom-resonant photon pairs come from degenerate mode (see Fig.~\ref{fig:filterandcavity}). In order to test how much light outside the Rubidium D$_1$ line can pass through our FADOF, we split the light equally between the two different polarization paths of the filter setup by means of a half-wave plate put before the beam displacer (HWP~1 in Fig.~\ref{fig:opo}). A natural-abundance Rb vapor cell, with 10~Torr of  N$_2$ buffer gas and heated until it is opaque for resonant light, is inserted in one of the paths  after the filter.  The collisionally-broadened absorption from this cell blocks the entire FADOF transmission window, allowing us to compare the arrival time histograms with and without the resonant component. 

The number of photons detected after passing through the hot Rb cell is comparable to the detector dark counts, meaning that most of the filtered light is at the chosen frequency $\omega_0$.  We define the spectral purity $\SP$ of the FADOF as $\SP \equiv 1 - c_{HC}/c_{F}$, where $c_{HC}$ ($c_{F}$) is the number of photon pairs which were recorded within a coincidence windows of 50~ns in the path with (without) the hot cell. Considering raw coincidences (no background subtraction), we obtain $\SP = 0.98$, meaning that the filtered signal is remarkably pure, as only the $2\%$ of the recorded pairs are out of the filter spectrum.  This $2\%$ agrees with measurements of the polarization extinction ratio with the FADOF off, i.e., it is due to technical limitations of the polarization optics and could in principle be improved. Knowing that $98\%$ of the photon pairs transmitted through the filter within the Rb resonance come from the degenerate cavity mode (due to filter spectrum), we conclude that  $96\%$ of the pairs exiting the filter come from the degenerate mode.

\subsection{Interference of biphoton amplitudes from distinct sources}
\label{sec:BiphotonInterference}
An SPDC process naturally generates a state of the form
\begin{equation}
\ket{\psi} \propto \ket{0} + \sum_{k_s,k_i}  f(k_s,k_i) a^\dagger_{k_s}a^\dagger_{k_i} \ket{0} + \ldots,
%\ket{\psi} \propto \ket{0} + \int dt\, dt'\, \psi(t,t') a^\dagger(t) a^\dagger(t') \ket{0} + \ldots,
\end{equation}
where $k_s$ and $k_i$ index the modes of the signal and idler fields, respectively, and $f$ is a complex-valued function analogous to the wave function encountered in non-relativistic quantum mechanics.  Following this analogy, one might expect that observations can only indicate $f$ up to an unobservable global phase.  A bit of reflection shows, however, that even the global phase of $f$ is observable, so long as the two-photon part of the wave-function exists in superposition with other parts, most importantly the zero-photon contribution $\ket{0}$.  

Consider the simplest scenario, in which signal and idler are the same mode, and differ at  most by their times of arrival $t$, $t'$.  This is in fact the case when we consider the filtered output of the CESPDC source described above.  The SPDC state becomes
\begin{equation}
\ket{\psi} \propto \ket{0} + \int dt\, dt'\, \psi(t,t') a^\dagger(t) a^\dagger(t') \ket{0} + \ldots,
\end{equation}
where $\psi$ now plays the role of  $f$. 
The phase of $\psi$ can be made visible by interfering the state against another state containing both a zero-photon and a two-photon component.  A natural candidate is the continuous-wave coherent state with amplitude $\alpha$:
\begin{equation}
\ket{\alpha} \propto \ket{0} + \alpha \int dt\, a^\dagger(t) \ket{0} + \frac{\alpha^2}{{2}} \int dt\, dt'\, a^\dagger(t) a^\dagger(t') \ket{0} + \ldots
%\ket{\psi} \propto \ket{0} + \int dt\, dt'\, \psi(t,t') a^\dagger(t) a^\dagger(t') \ket{0} + \ldots,
\end{equation}

\begin{figure}[t]
\centering
   \includegraphics[width=0.82\columnwidth]{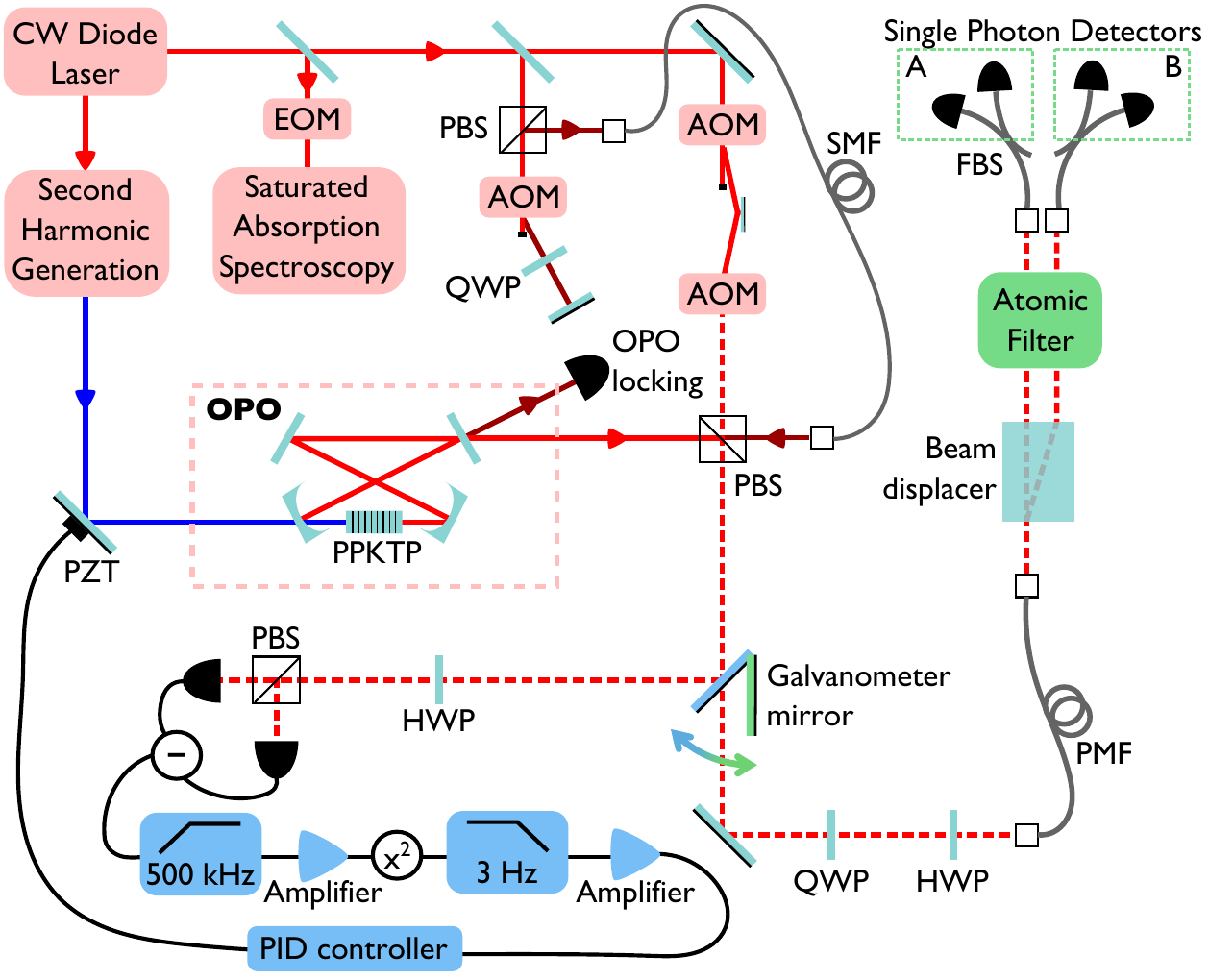}
   \caption{Experimental setup. AOM (EOM): acousto- \mbox{(electro-)} optic Modulator. PBS: polarizing beam splitter. QWP (HWP): quarter- (half-)wave plate. PZT: piezoelectric actuator. SMF: single mode fiber. PMF: polarization maintaining fiber. FBS: fiber beam splitter. }
   \label{fig:setup}
\end{figure}

We can imagine combining at a polarizing beamsplitter the states $\ket{\psi}$, with vertical polarization, and $\ket{\alpha}$, with horizontal polarization, to obtain a state 
%\begin{equation}
%\ket{\Psi} \equiv \ket{\psi}_V \otimes \ket{\alpha}_H
%\end{equation}
\begin{eqnarray}
\ket{\Psi} &\equiv& \ket{\psi}_V \otimes \ket{\alpha}_H \\ 
& = & \left[ 1 + \alpha \int dt\, a^\dagger_H(t) +  \int dt\, dt'\, \psi(t,t') a_V^\dagger(t) a_V^\dagger(t') \right. \nonumber \\ & & + \left.  \frac{\alpha^2}{2} \int dt\, dt'\, a_H^\dagger(t) a_H^\dagger(t') + \ldots \right]  \ket{0}_V \otimes \ket{0}_H 
%& = & \ket{0}_V \otimes \ket{0}_H + \alpha \int dt\, a^\dagger(t) \ket{0}_V \otimes \ket{0}_H 
\end{eqnarray}
with a two-photon component 
\begin{equation}
 \int dt\, dt'\, \left[ \psi(t,t') a_V^\dagger(t) a_V^\dagger(t') + \frac{\alpha^2}{2} a_H^\dagger(t) a_H^\dagger(t') \right]  \ket{0}_V \otimes \ket{0}_H.
 \end{equation}
Interference of these two terms can be obtained by coincidence detection in a basis that is neither $V$ nor $H$.  For example, the rate of detection of a photon pair, at times $t$ and $t'$, both in the state $\ket{\phi} \equiv (\exp[-i \phi] \ket{H} + \ket{V})/\sqrt{2}$, is found by projection onto $\ket{ \phi}^{\otimes 2}$ to give 
\begin{equation}
\label{eq:TwoPhotInterf}
R_\phi(t,t') \propto \left| \psi(t,t') + \frac{\alpha^2}{2} \exp[2 i \phi] \right|^2,
 \end{equation}
which clearly depends on the phase of $\psi$.  It should be noted that observing this interference requires both narrow-band photon pairs, so that the detection times can be resolved, and a stable phase relation between $\alpha$ and $\psi$.  To maintain this phase relationship, we use the OPO also as a phase-sensitive amplifier, a well-established technique from continuous-variable quantum optics.

The experimental setup is shown in Fig. \ref{fig:setup}.  A continuous-wave diode laser at 794.7~nm generates both the coherent reference beam and, after being amplified and doubled in frequency, a 397.4~nm pump beam for the OPO, described in~\cite{PredojevicPRA2008}, which generates a vertically-polarized (V) squeezed vacuum state via SPDC in a periodically poled KTP crystal.  The cavity length is actively stabilized with a Pound-Drever-Hall lock, to keep one longitudinal V mode resonant at the laser frequency.  The locking beam is H polarized, counter-propagating, and shifted in frequency by an acousto-optic modulator (AOM), to match the frequency of an H-polarized mode.  The AOM RF power is chopped and the detectors are electronically gated: coincidence data are acquired only when the locking light is off.  With these measures, the contribution of locking light \ntext{to the accidental coincidences background} is minimised. 

The V-polarized squeezed vacuum is combined with the H-polarized coherent reference at a polarizing beamsplitter to generate a beam with co-propagating squeezed and reference components.  A polarization transformation, chosen so that $\ket{\phi} \equiv (\exp[-i \phi] \ket{H} + \ket{V})/\sqrt{2}$ arrives to one detector, is implemented with a quarter- and a half-waveplate, before coupling into a polarization-maintaining fiber.   
  
At the fiber output, the two polarization components are filtered 
%separated into parallel beams by a calcite beam displacer and passed through
with the two-polarization FADOF described in Section \ref{sec:Experiment}, in order to isolate the squeezed vacuum and block with high efficiency the hundreds of non-degenerate frequency modes generated by the OPO. The maximum transmission frequency of this filter is located \ntext{at 2.7~GHz to the red} of the center of the rubidium D${}_1$ line,  and the laser frequency is stabilised at this particular frequency by using an integrated electro-optic modulator to add sidebands to the laser prior to the saturated absorption spectroscopy.  
%Each filtered beam is then coupled into a single-mode fiber and split with a 50/50 fiber beam splitter to a pair of single-photon counting avalanche photo diodes.  A time-of-flight recorder time-stamps each arrival and correlations are computed on a PC.  
\begin{figure}[bt]
   \centering
   \includegraphics[width=\columnwidth]{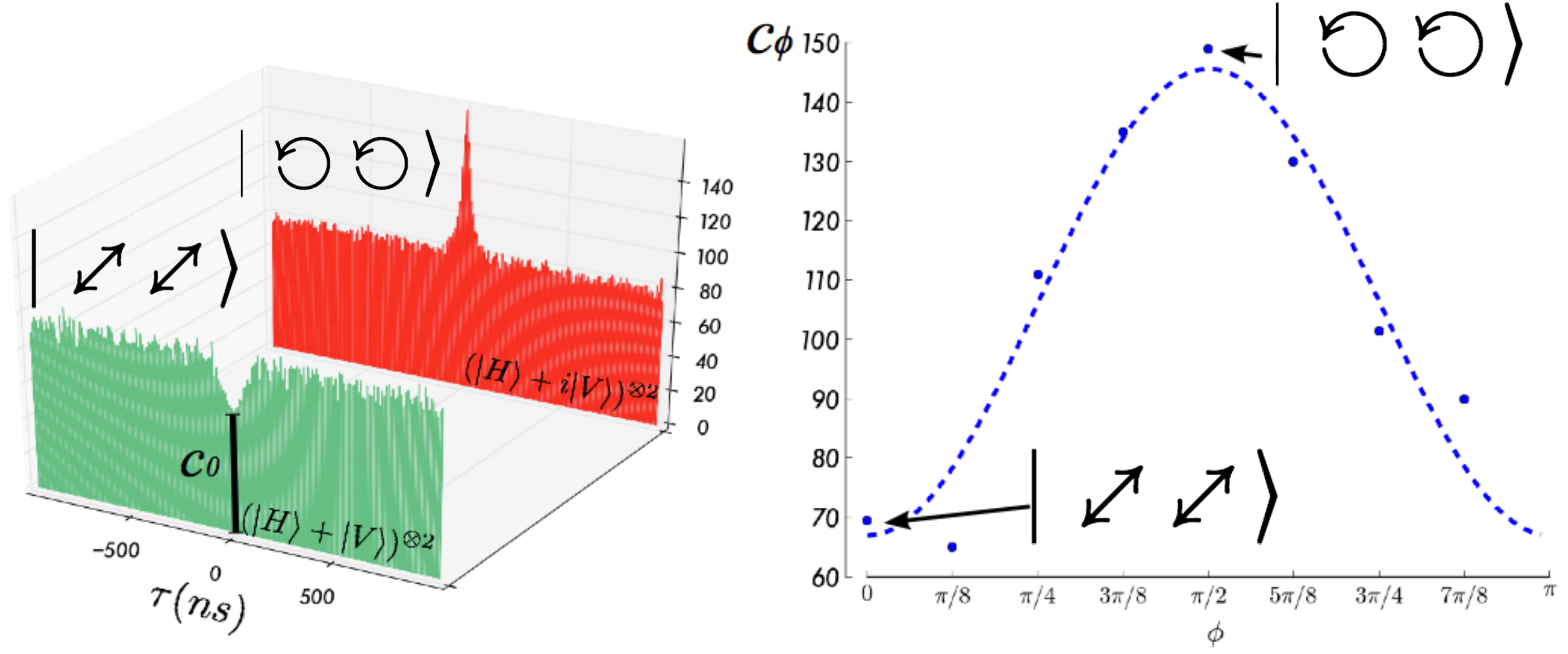}
   \caption{Time-resolved interference of the two-photon amplitude from CESPDC against the two-photon amplitude from a coherent state.  (left) arrival-time distribution, photon pair rate as a function of time difference $\tau$, for photons pairs either both diagonally-polarized or both circularly-polarized.  A $\tau$-independent coherent state contribution interferes with a CESPDC contribution that has a double-exponential form centred on $\tau = 0$, with a FWHM width of 26~ns.  As seen in the two histograms, the interference can be either constructive or destructive.  (right) height of the interference peak versus the phase $\phi$ appearing in the two-photon state $(\exp[-i \phi] \ket{H} + \ket{V})/\sqrt{2}$.  As expected for a two-photon interference, and strikingly different from single-photon interference, the period of oscillation is $\pi$.   }
   \label{fig:visibility}
\end{figure}

The relative phase $\relphase$ between the coherent and the squeezed beam is stabilized by a quantum noise lock:  One Stokes component is detected with a balanced polarimeter, and the noise power in a 3 Hz bandwidth above 500 kHz is computed analogically using a multiplier circuit.  This signal is fed back by a servo loop to a piezo-electric actuator on a mirror in the pump path, to stabilize the pump phase by a side-of-fringe lock.  A galvanometer mirror is used to switch between the single-photon counting and stabilisation setups at a frequency of $\sim$100 Hz.  The reference beam power is increased during the stabilization part of the cycle, to reach the shot-noise-limited regime optimal for detection of the squeezing and operation of the noise lock. Two cascaded AOMs, whose RF power is chopped synchronously with the galvanometer mirror, modulate the coherent reference beam power, so that it has high power when the light is entering the stabilisation setup and low power when the photon counting part is active. The system can maintain a fixed  $\relphase$  over several hours.

Results are shown in Fig. \ref{fig:visibility}, and clearly show both constructive and destructive interference of the two-photon wave-function against the coherent state.

\begin{figure}[t]
   \centering
\includegraphics[width=0.80\columnwidth]{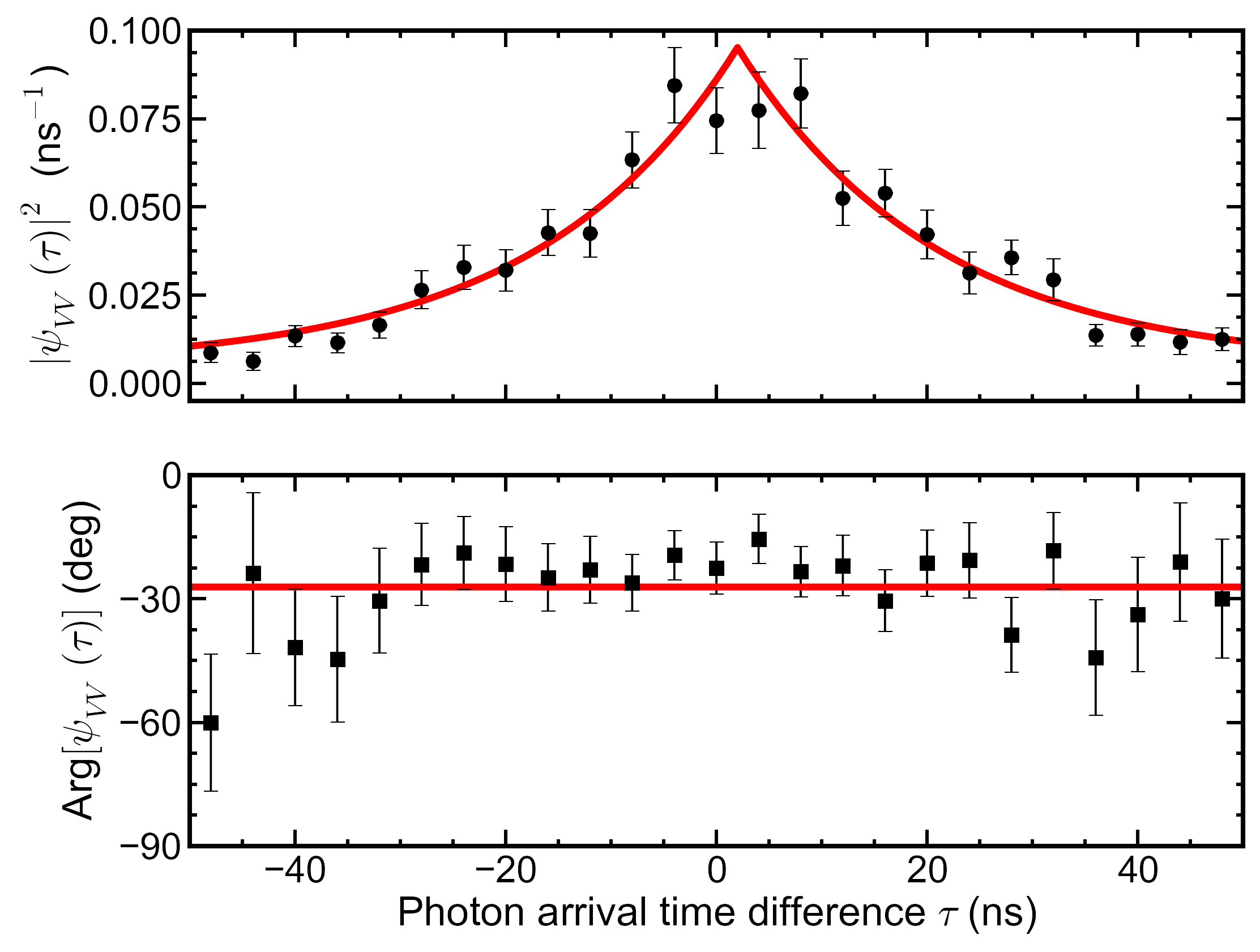}
   \caption{Squared amplitude (above) and phase (below) of the reconstructed two-photon wave function for the squeezed vacuum state. \ntext{The solid line shows the predicted, double exponential amplitude describing an ideal squeezed vacuum state from our OPO with an independently-measured 8.1~MHz bandwidth, {with the amplitude and the offset fitted to the data}. {Error bars show $\pm 1\sigma$ statistical uncertainty assuming Poisson statistics. %and using propagation of error through Eqs. \eqref{eq:psiV} and~\eqref{eq:gamma}.   
 }
    } } \label{fig:TPWF}
\end{figure}

\subsection{Full reconstruction of the biphoton wave-function}
\label{sec:Reconstruction}

The form of Eq. (\ref{eq:TwoPhotInterf}) suggests a method to measure the two-photon wave-function, including not only its amplitude, but also its phase.  If the term $\frac{\alpha^2}{2} \exp[2 i \phi]$ is under our control, it should be possible by setting this term and measuring the resulting coincidence rates, to infer the complex value of $\psi(t,t')$.  This idea was made precise in \cite{BeduiniPRL2014}, and the above setup was used to reconstruct the biphoton wave-function shown in Fig. \ref{fig:TPWF}.  

The results are consistent with a double-exponential amplitude with 26~ns full-width at half-maximum (FWHM), as expected for a squeezed vacuum state from an OPO with the 8.1~MHz FWHM bandwidth independently-measured on our system.   {The phase of  $\psi_{VV}^{(\lambda)}$, consistent with a non-zero constant value, is reconstructed with a statistical uncertainty that decreases with increasing  $|\psi_{VV}^{(\lambda)}|$, reaching $\sigma_\phi \approx \pm 6$ degree near $\tau=0$.  } A constant phase is expected for an ideal OPO, while a phase defect could signal cavity or crystal imperfections \cite{KuzucuPRA2008,ODonnellPRL2009}.  The phase offset is tunable via the side-of-fringe lock that sets the relative phase of the squeezed vacuum and reference, and is another indication of interference at the two-photon level.

\section{Generation of spectrally-pure, atom-resonant \noon states }

\btext{In this section we describe a series of experiments to generate atom-tuned photon pairs from Type-II SPDC.  Subsection \ref{sec:NooNStates} presents a motivation in terms of \noon states and their interest for quantum-enhanced sensing, subsection \ref{sec:TypeII} describes the SPDC source and characterization of the generated states, subsection \ref{sec:InducedDichroFilter} describes the atomic filter, subsection \ref{sec:PurityII} describes measurements of the spectral purity achieved,  and subsection  \ref{sec:QESNooN} describes the application of atom-tuned \noon states to quantum-enhanced sensing of magnetic fields using an atomic ensemble as a sensor.}   

%We now describe the generation of atom-tuned photon pairs of orthogonal polarization, with one photon $H$-polarized and one $V$-polarized, but indistinguishable in all other degrees of freedom. 

\subsection{NooN states}
\label{sec:NooNStates}
We now describe the generation of atom-tuned photon pairs of orthogonal polarization, with one photon $H$-polarized and one $V$-polarized, but indistinguishable in all other degrees of freedom. 
A state of one $H$ state can be written
\begin{eqnarray}
\ket{\psi} &=& a^\dagger_H a^\dagger_V \ket{0} = \frac{1}{2} \left( a^\dagger_L a^\dagger_L - a^\dagger_R a^\dagger_R \right) \ket{0} 
\end{eqnarray}
where the circular polarization modes are defined by $a_L = (a_H + i a_V)/\sqrt{2}$, $a_R = (a_H - i a_V)/\sqrt{2}$.  When written in terms of the photon numbers $n_L$, and $n_R$ in the $L, R$ modes respectively, and using the notation $\ket{n_L, n_R}_{L,R}$, this state is
\begin{eqnarray}
\ket{\psi} &=& \frac{1}{\sqrt{2}} \left( \ket{N,0}_{L,R} + \ket{0,N}_{L,R} \right)
\end{eqnarray}
with $N=2$, an example of a ``NooN'' state, named for the letters that appear in the kets $\ket{\cdot}$. 

Consider a linear polarization interferometer that operates by Faraday rotation.  This imposes a differential phase $\phi$ between the $L$ and $R$ parts of the state, transforming the NooN state as
\begin{eqnarray}
\frac{1}{\sqrt{2}} \left( \ket{N,0}_{L,R} + \ket{0,N}_{L,R} \right) & \rightarrow & 
\frac{1}{\sqrt{2}} \left( \ket{N,0}_{L,R} + e^{i N \phi} \ket{0,N}_{L,R} \right).
\end{eqnarray}
Note that the phase acts $N$ times on the second part of the state, because it contains $N$ photons.  This implies that any signal derived from this state must vary with $N \phi$, implying an $N$-fold acceleration of any interferometric signal.  If we consider that $\phi$ is an unknown phase, this accelerated interference implies an $N$-fold increase in the Fisher information \cite{%PezzeARX2014,
 PezzeISEF2014}, allowing estimation of $\phi$ with uncertainty $\delta \phi = 1/N$, improving upon the standard quantum limit of $\delta \phi = 1/\sqrt{N}$, the best obtainable with non-entangled states. 

A major motivation of this work is to test the suitability of entangled states for quantum enhanced sensing with atoms.  In atomic media, interferometric phase shifts are necessarily accompanied by absorption, implying deposition of energy in the probed medium.  Absorption also degrades any quantum advantage, as  described by recent theory \cite{DornerPRL2009,EscherNP2011}.  To further complicate matters, in real media the phase shift and absorption may depend on the same unknown quantity. %, a feature not yet considered in theory
In a trade-off of rotation strength versus transparency, we employ a \noon state in a \SI{7}{MHz} spectral window detuned four Doppler widths from the nearest $^{85}$Rb resonance.  We generate this, as above, with a CESPDC source and an ultra-narrow atom-based filter.

% This state is of interest in quantum enhanced sensing, because in the ideal scenario of lossless linear interferometry, it achieves the so-called Heisenberg limit, extracting the maximum information that is possible with $N$ photons \cite{MitchellN2004}.  Our motivation here is to use this state for readout of an atomic sensor, to demonstrate and test the principles of quantum metrology when entangled photons are used for quantum enhanced sensing of material systems.  

\subsection{Type-II CESPDC source}
\label{sec:TypeII}

\begin{figure}[t]
\centering
\includegraphics[width=8.3cm]{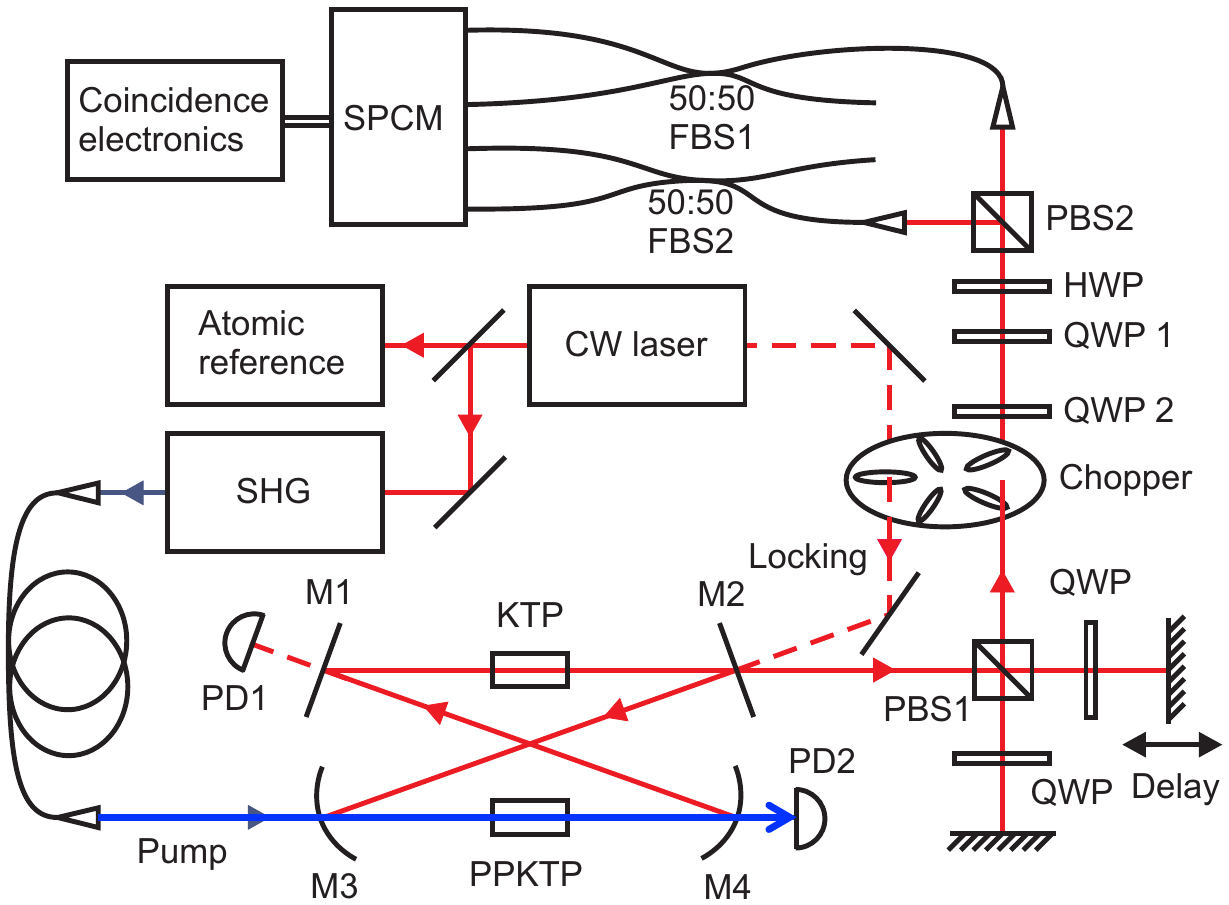}
\caption{Type-II CESPDC source: PPKTP, phase-matched nonlinear crystal;
KTP, compensating crystal; M1-4, cavity mirrors; PBS, polarizing
beam splitter; HWP, half wave plate; QWP, quarter wave plate; SMF,
single-mode fiber; PD, photodiode.}
\label{fig:TypeIISetup}
\end{figure}

%We use type-II \cespdc~ %\cite{Bao2008, Scholz2009, Scholz2009b}
% \cite{Kuklewicz2006, Wolfgramm2008, Wolfgramm2010} to produce $N=2$ \noon states. A 795 nm laser, stabilized to $\omNOON$, the frequency of the $5^{2}{\rm S}_{1/2} {\rm F}$=$2 \rightarrow 5^{2}{\rm P}_{1/2} {\rm F}'$=$1$ transition of the $D_1$ line of $^{87}$Rb, is both used to stabilize the cavity length, and is frequency-doubled to pump the SPDC process. In the down-conversion process, the cavity resonantly enhances degenerate down-conversion into a TEM$_{00}$ mode with bandwidth $7$ MHz around $\omNOON$. Non-degenerate emission, separated by at least the cavity's 490 MHz free spectral range, is efficiently removed by filtering. The apparatus is described in detail in references
%\cite{Wolfgramm2008,Wolfgramm2010}. By using a low pump power, we
%ensure that the emitted photons are in a two-photon state
%$\rhoin$, ideally the \noon state $|1\rangle_H|1\rangle_V \propto |2\rangle_L|0\rangle_R - |0\rangle_L|2\rangle_R $, with negligible four-photon component. $H,V,L,R$ indicate horizontal, vertical, left circular (or $\sigma_+$) and right circular (or $\sigma_-$) polarizations, respectively.

As laser source we use a continuous-wave (CW) diode laser, stabilized to the
D$_1$ transition of atomic rubidium at 795~nm and then frequency
doubled to generate a 397.5~nm pump that is passed through a
mode-cleaning single-mode fiber and then focused into the center of a
20 mm-long periodically-poled KTiOPO$_4$ \mbox{(PPKTP)} crystal in a
cavity, forming the OPO (Fig.\ \ref{fig:TypeIISetup}). A pump beam waist
of 30~\SI{}{\micro m} is achieved with a telescope. This beam waist was
chosen to be larger than the optimum for degenerate down-conversion
according to Boyd and Kleinman \cite{BoydJAP1968} in order to reduce
possible effects of thermal lensing \cite{LeTargatOC2005} and
gray-tracking \cite{BoulangerAPL1994}. The crystal is poled for type-II
degenerate down-conversion, and produces orthogonally-polarized
signal and idler photons.  Due to crystal birefringence, these
photons experience temporal walk-off that would, if un-compensated,
render the photons temporally distinguishable.  A second KTP crystal
of the same length and crystal cut, but not phase-matched and
rotated about the beam direction by 90$^{\circ}$, is added to the
long arm of the cavity in order to introduce a second walk-off equal
in magnitude but opposite in sign \cite{KuklewiczPRL2006}.

The ring cavity is formed by two flat mirrors (M1, M2) and two
concave mirrors (M3, M4) with a radius of curvature of 100~mm. The
effective cavity length of 610~mm corresponds to a FSR of 490~MHz. This geometry provides a beam waist of
42~\SI{}{\micro m} for the resonant down-converted beam at the center of
the crystal, which matches the 30~\SI{}{\micro m} pump beam waist. 
Cavity length is controlled by a piezoelectric
transducer on mirror M1. The output coupler M2 has a reflectivity of
93\% at 795~nm. All other cavity mirrors are highly reflecting
(R~$>$~99.9\%) at 795~nm and highly transmitting at 397.5~nm
(R~$<$~3\%) resulting in a single-pass through the nonlinear crystal
for the blue pump beam. The crystal end faces are AR coated for
397.5~nm and 795~nm. The measured cavity finesse of 70 results in a
cavity linewidth of 7~MHz.

While the walk-off per round trip is compensated by the KTP crystal,
there is an uncompensated walk-off of in average half a
crystal-length, because of the different positions inside the PPKTP,
where a photon pair could be generated. This leads to a remaining
temporal distinguishability at the output of the cavity that is
completely removed by delaying the horizontally polarized photon of
each pair with a Michelson-geometry compensator: a polarizing beam
splitter, retro-reflecting mirrors, and quarter wave-plates set to
45$^{\circ}$ introduce an adjustable delay while preserving spatial
mode overlap. After recombination the pairs are sent through a half
wave plate (HWP2) that together with PBS2 determines the measurement
basis. Both output ports of PBS2 are coupled into single-mode fibers
(SMF) connected to single photon counting modules (Perkin Elmer
SPCM-AQ4C). The pulse events are registered and processed by
coincidence electronics (FAST ComTec P7888) with a resolution of
1~ns.

The OPO cavity is actively stabilized by injecting an auxiliary
beam, derived from the \btext{same} diode laser, into the cavity via the output
coupler (M2). This light is detected in transmission by a photodiode
(PD1). Frequency modulation at 20~MHz\btext{, applied via the laser diode current,} is used to lock to the peak of
the cavity transmission. To eliminate the background noise caused by
this auxiliary beam and to protect the SPCMs, the locking and
measuring intervals are alternated using a mechanical chopper at a
frequency of about 80~Hz with a duty cycle of 24\%.

A general polarization analyzer, consisting of a quarter wave
plate (QWP1) followed by a half wave plate (HWP) and a polarizing
beam splitter (PBS2) is used to determine the measurement basis as
shown in Fig.\ \ref{fig:TypeIISetup}. To generate a NOON state in the
$H$/$V$ basis another quarter wave plate (QWP2) can be added. The
two output ports of PBS2 are coupled to single-mode fibers and
split with 50:50 fiber beam splitters. The four outputs are
connected to a set of single photon counting modules (Perkin Elmer
SPCM-AQ4C). Time-stamping was performed by coincidence electronics
with a resolution of 2~ns. By considering a time window of 150~ns,
which is longer than the coherence time of each individual photon,
we can evaluate the coincidences between any two of the four
channels.

%\begin{figure}[t]
%\centering
%\includegraphics[width=7cm]{../images/1quality.eps}
%\caption{Experimental Setup. SHG, second harmonic generation
%cavity; PPKTP, phase-matched nonlinear crystal; KTP, compensating
%crystal; M1-4, cavity mirrors; PBS, polarizing beam splitter; HWP,
%half wave plate; QWP, quarter wave plate; SMF, single-mode fiber;
%PD, photodiode; FBS, fiber beam splitter; SPCM, single photon
%counting module.\label{fig:TypeIISetup}}
%\end{figure}

\subsubsection{Characterization of the \noon state}

\begin{figure}[t]
\centering
\includegraphics[width=0.9\textwidth]{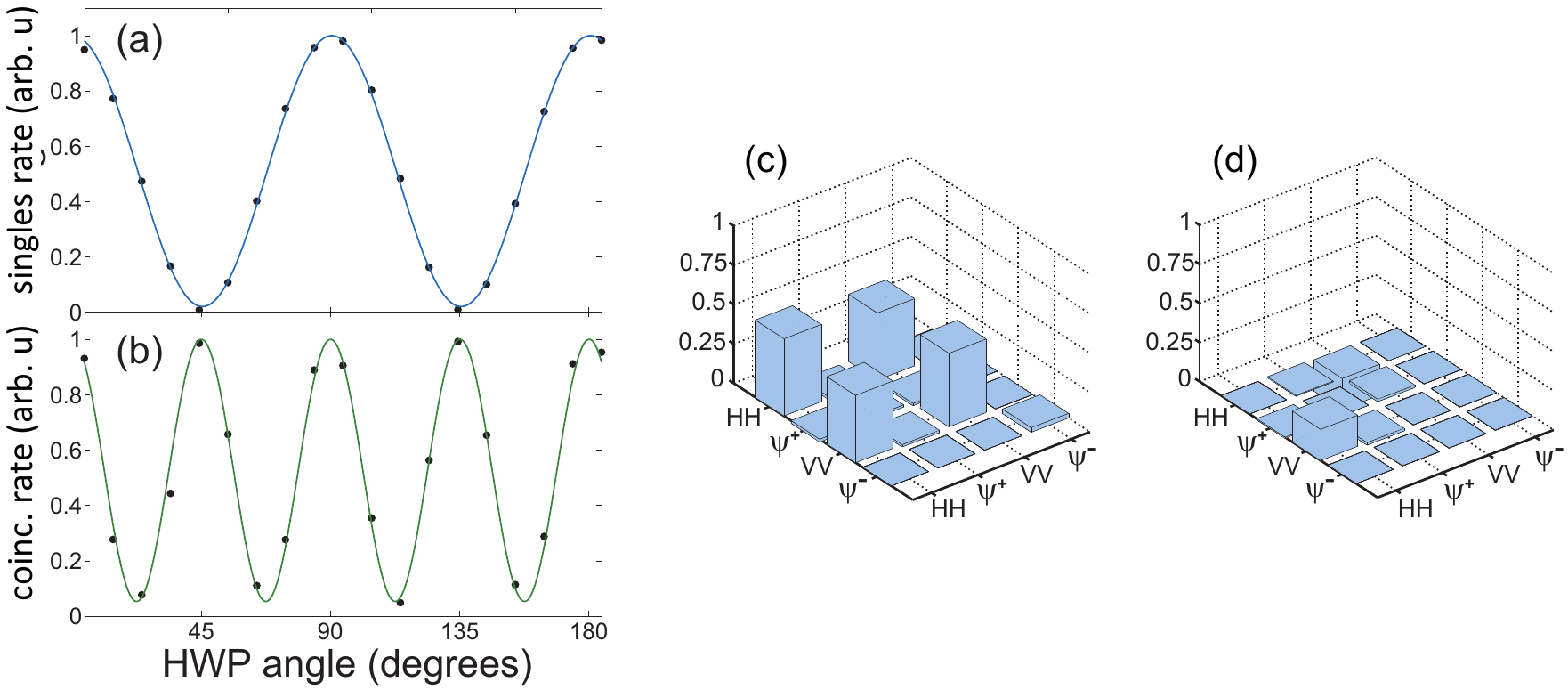}
%\hspace{\fill}
%\includegraphics[width=0.60\textwidth]{../images/4quality-eps-converted-to.pdf}
\caption{Multi-photon interference and state characterization.  Left panels:
resolution of one- and two-photons states under rotations.  (a) Single-photon
rotation measurement showing ordinary oscillations with a period of 90$^\circ$
in the HWP angle.   In this measurement
only the $H$ polarized part of the pair-photon state was sent to
the analyzer. Vertical axis shows normalized singles
rate at the transmitted port of PBS2. (b) High-visibility super-resolving phase measurement. Normalized
coincidence detection between reflected and transmitted port of
PBS2 for a \noon state input. The shorter period of the coincidence
counts oscillations indicates super-resolution. 
Right panels:  results of quantum state tomography to characterize the \noon state.  
(c) Real and (d) imaginary part of the polarization
density matrix of the pair-photon state transformed to a
two-photon NOON state in the $H$, $V$ basis, showing a 
nearly equal superposition of $\ket{HH}$ and $\ket{VV}$.}
\label{img:NOON_states}
\label{img:Super-Resolution}
\end{figure}

Rotating the HWP before the detection setup allows us to demonstrate the
greater resolution available with the \noon state, as shown in Fig.~\ref{img:NOON_states} (left).
  First, sending just a single 
polarization and detecting the rate of single-photon arrivals, we observe the expected
oscillations with a period, in HWP angle, of 90$^\circ$.  Then, sending \noon states and 
detecting in coincidence, we observe a two-fold reduction in the oscillation period simultaneous
with high visibility, due to the two-photon coherence of the \noon state. 
The sinusoidal fit function of the coincidences shows a  visibility of 90\%.

%\section{High quality NOON state}
The achieved high visibility of the state is the requirement for a
high-fidelity NOON state. We introduce another quarter wave plate
(QWP2) before the analyzing part of the setup to create a
two-photon NOON state in the $H$/$V$ basis, which can be written
$1/\sqrt{2}(|H_1,H_2\rangle+e^{2i\phi} |V_1,V_2\rangle)$. Since the
output state of the cavity $|HV\rangle$ is already a NOON state in
the circular basis $|HV\rangle=i/\sqrt{2}(|L_1,L_2\rangle-
|R_1,R_2\rangle)$, this state can be transferred into a NOON state
in the $H$/$V$ basis by sending it through an additional quarter
wave plate at 45 degrees.

We also use quantum state tomography, as in \cite{AdamsonPRL2007}, to measure
the polarization density matrix of the \noon state \cite{WolfgrammJOSAB2010}. 
In Fig.~\ref{img:NOON_states} (right) real and imaginary parts of the
reconstructed density matrix of a NOON state are displayed. The
coherence of the state is partly imaginary leading to a phase of
$2\phi=0.20$ between $HH$ and $VV$ components
(Fig.~\ref{img:NOON_states}(b)), which is however of no
importance in the following. The fidelity of this state with the
corresponding ideal two-photon NOON state
$1/\sqrt{2}(|H_1,H_2\rangle+e^{2i\phi} |V_1,V_2\rangle)$ is 99\%,
making the state suitable for applications such as
phase estimation \cite{MitchellN2004}. 
%To demonstrate this ability,
%we performed a super-resolving phase experiment. After passing the
%NOON state, for this experiment in the circular basis (without
%QWP1 and QWP2), through the HWP, the coincidence counts between
%the output ports of PBS2 for different HWP settings were recorded.
%In Fig.~\ref{img:Super-Resolution} the interference fringes of
%the coincidences are displayed together with single counts from a
%measurement in which one polarization of the cavity output was
%blocked.

%\begin{figure}[h]
%\centering
%\includegraphics[width=5cm]{../images/5quality.eps}
%\caption{(a) Standard phase measurement. Normalized singles
%detection at the transmitted port of PBS2. In this measurement
%only the $H$ polarized part of the pair-photon state was sent to
%the analyzer. (b) Super-resolving phase measurement. Normalized
%coincidence detection between reflected and transmitted port of
%PBS2 for a NOON state input. The shorter period of the coincidence
%counts oscillations indicates
%super-resolution.\label{img:Super-Resolution}}
%\end{figure}

\begin{figure}[htb]
\centering
\includegraphics[width=0.75 \columnwidth]{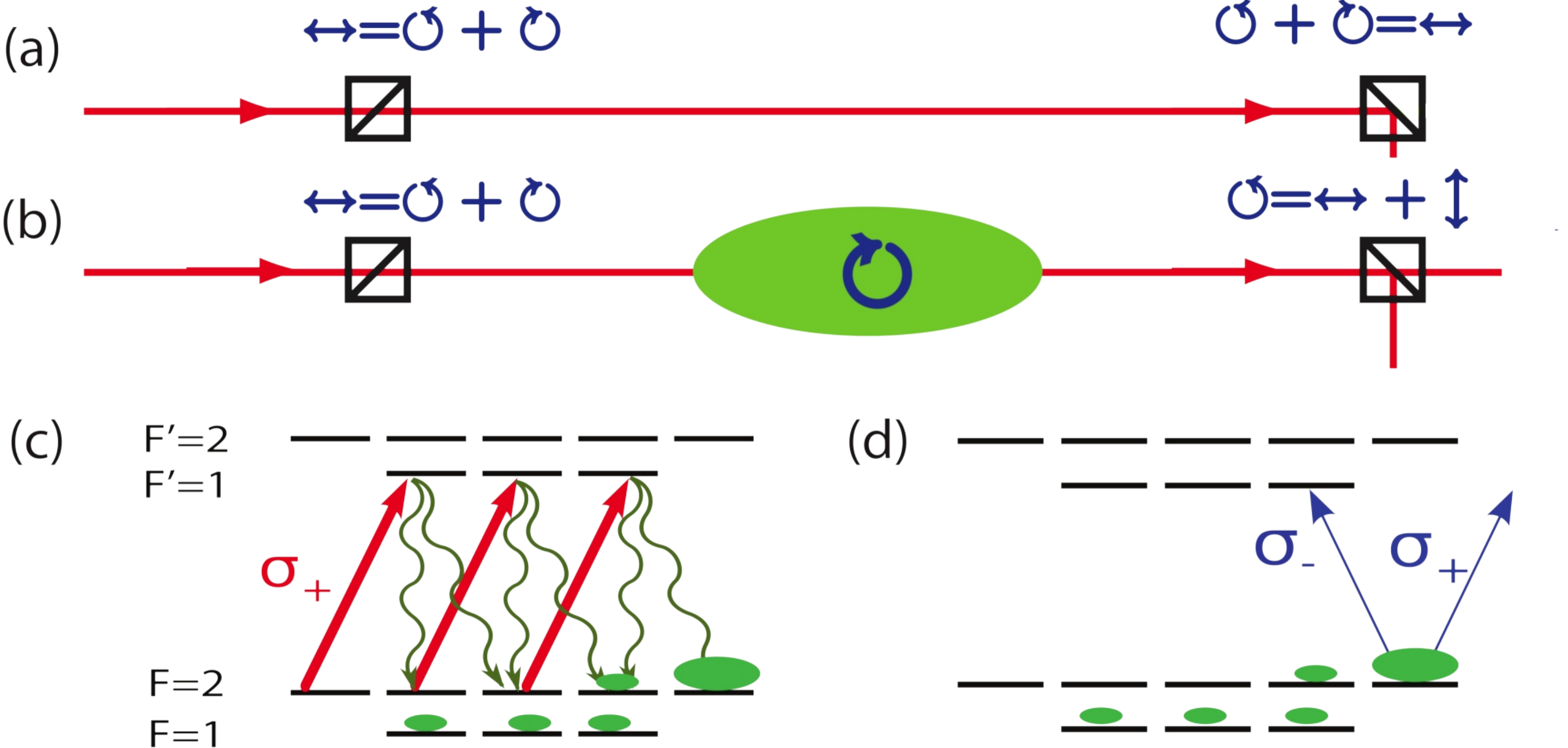}
 \label{fig:DichroFilterIdea}
\caption{Principle of the circular dichroism filter.  (a) non-resonant photons are blocked, either by the first or by the second PBS.  (b) resonant photons are linearly polarized after the first PBS, may become circularly polarized by interaction with the circularly dichroic atomic medium, and then have a chance of passing the second PBS.  In this way, atom-resonant light can pass the filter.  In the ideal case of a perfect circular absorber, the transmission would still only be 25\%.  (c) optical pumping, shown here on the D$_1$ line, moves atomic population toward the states $F=2, m_F = 2$ and $F=2, m_F = 1$.  (d) from these states, the atoms cannot absorb more $\sigma_+$ light on the $F=2 \rightarrow F'=1$ transition, creating a strong circular dichroism at this frequency.        }
\end{figure}

%\begin{figure}[htb]
%\centering
%\includegraphics[width=\columnwidth]{../images/Cere_105202_fig2.pdf}
% \label{fig:DichroFilterSetup}
%\caption{Setup.}
%\end{figure}

%\cite{ZhuangCOL2014}

\begin{figure}[htb]
\centering
\includegraphics[width=0.49 \columnwidth]{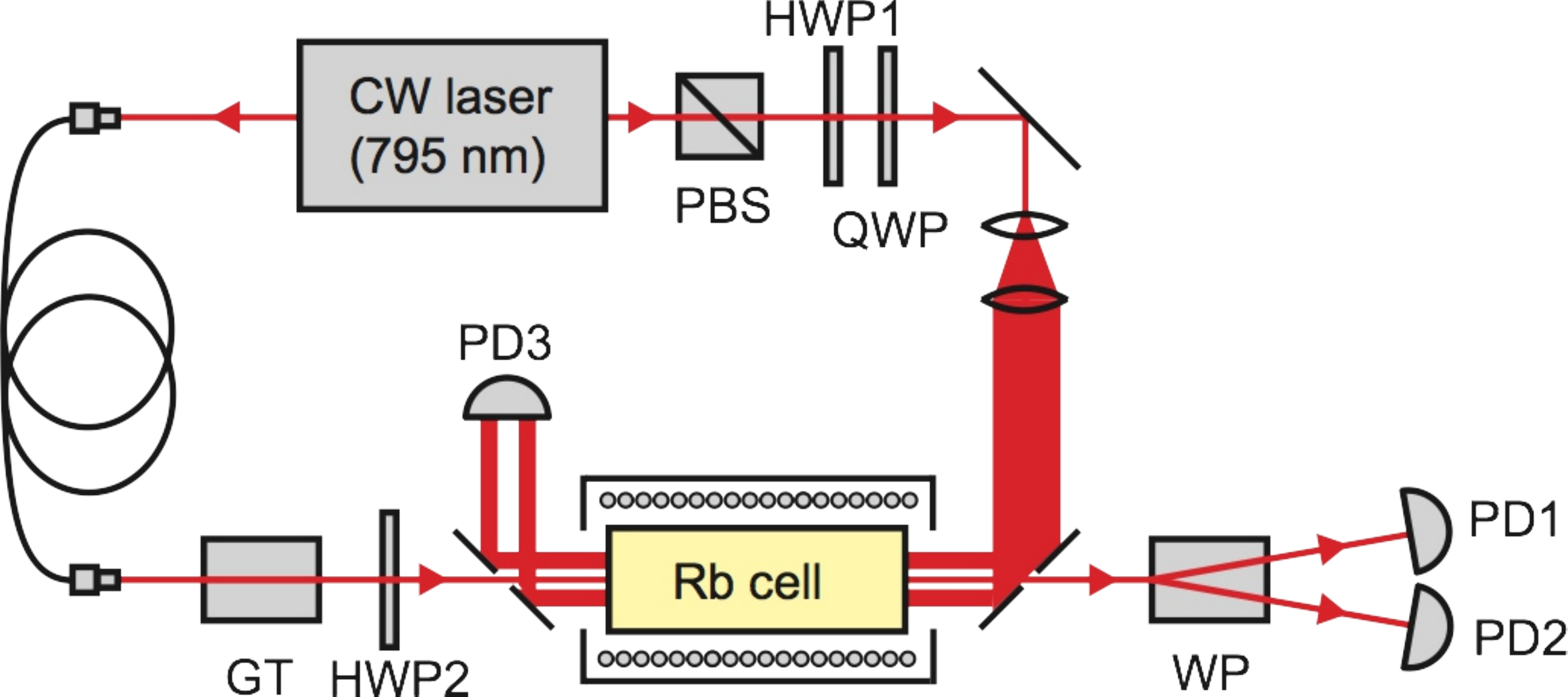}
\hspace{\fill}
\includegraphics[width=0.49 \columnwidth]{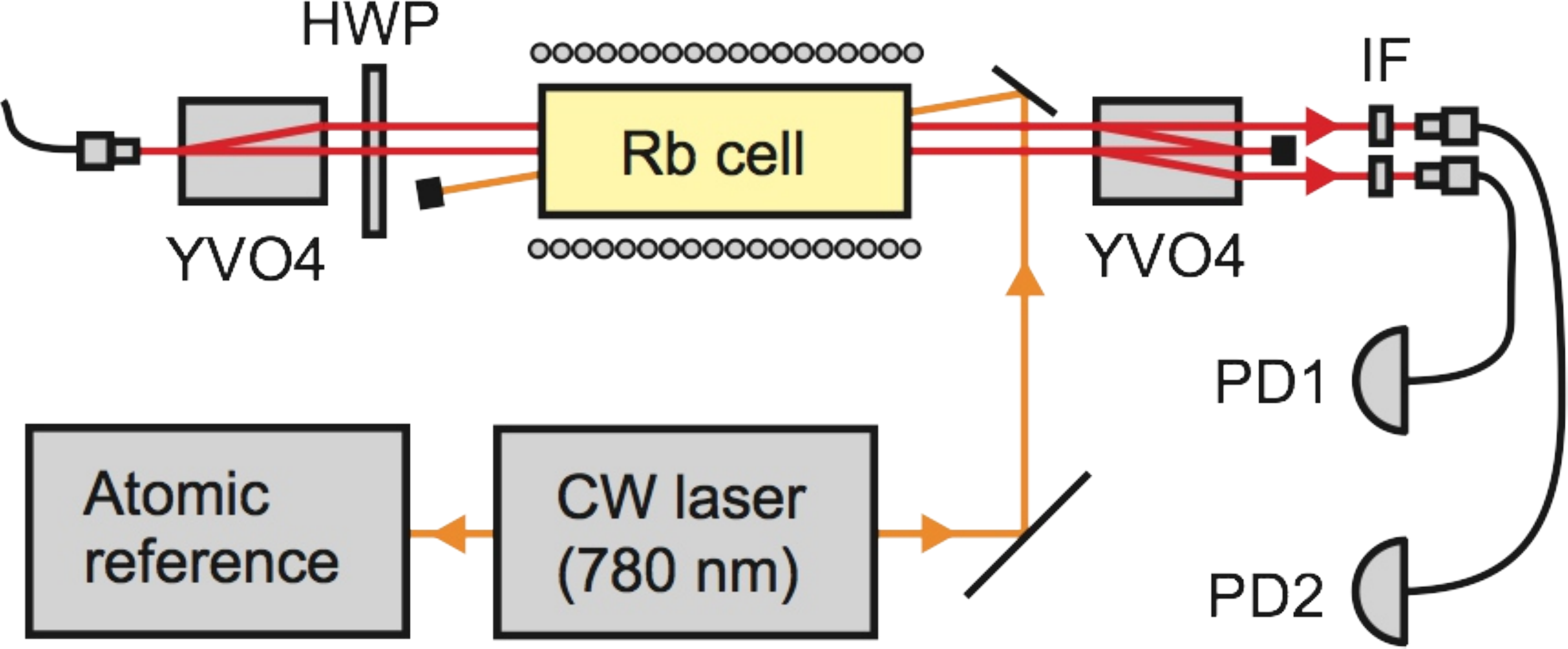}
\caption{Two arrangements for induced dichroism filtering.  left: pumping on the D$_1$ transition with geometric separation of the pump and probe beams, as in \cite{CereOL2009}.  By combining the pump and probe beams using pierced mirrors, the pump beam propagates in a ring-shaped beam, coaxial with the probe beam, to achieve optical pumping of the atoms before they reach the probe beam, but without geometrical overlapping that beam. right:  pumping on the D$_2$ transition at 780 nm with overlapped beams, as in \cite{WolfgrammPRL2011}.  The wavelength difference between the pump and probe allows pump light to be filtered using commercial interference filters (IF), and the use of YVO$_4$ walkoff crystals as polarizers allows both polarizations to be filtered and collected.  In both arrangements, the pump and probe are counter-propagating, giving rise to sub-Doppler resonances.   }
 \label{fig:DichroSetups}
\end{figure}

\begin{figure}[htb]
\centering
\includegraphics[width=0.9 \columnwidth]{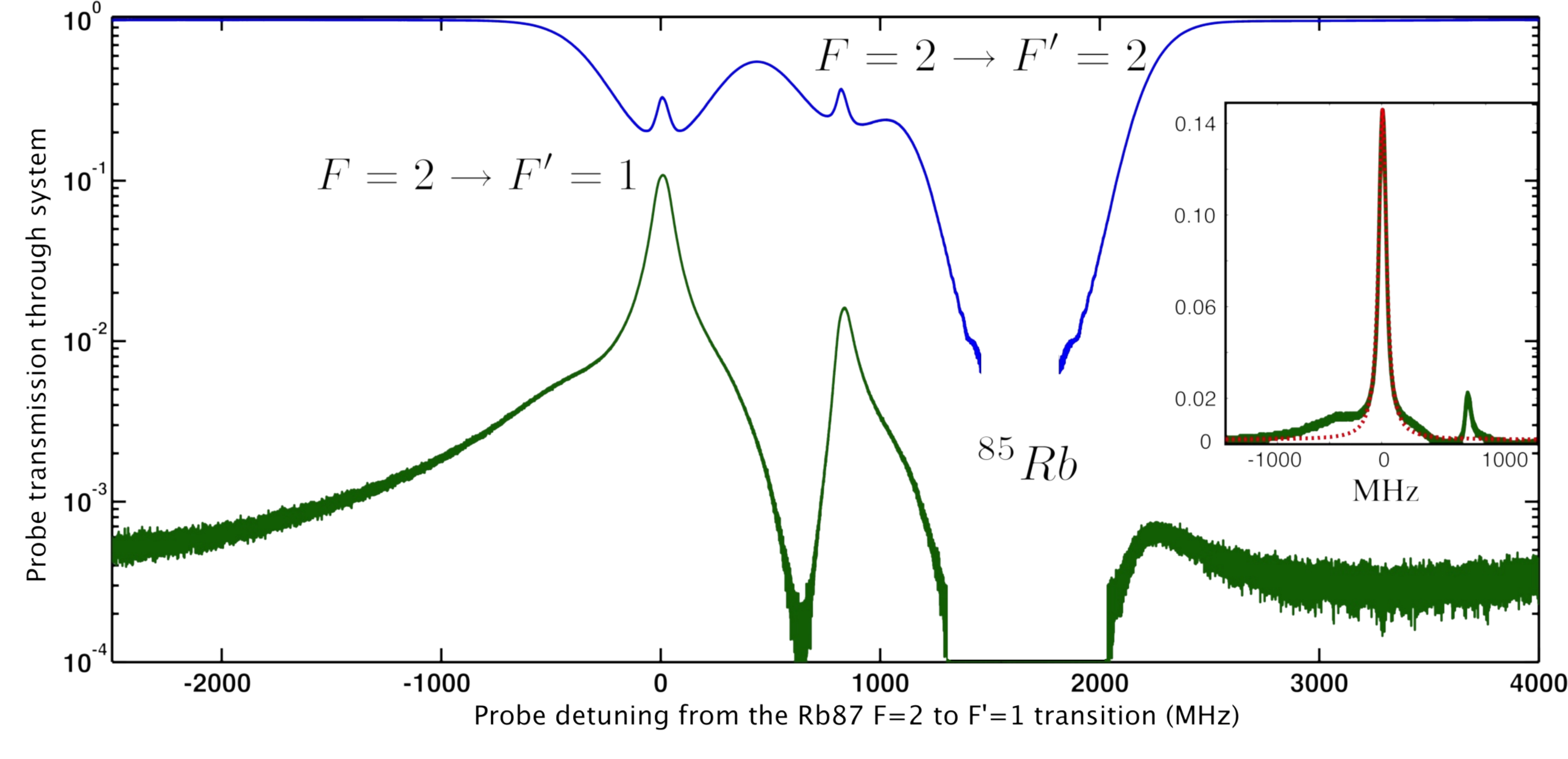}
 \label{fig:DichroFilterTrans}
\caption{Transmission of the circular dichroism filter, obtained with the configuration of Fig. \ref{fig:DichroSetups} (left).  Blank areas around 1500 MHz indicate regions where the transmission was unmeasurably low, due to the strong extinction of the $^{85}$Rb transition. }
\end{figure}

%\begin{figure}[htb]
%\centering
%\includegraphics[width=0.95 \columnwidth]{../images/WolfgrammMagSpectrum.pdf}
% \label{fig:DichroFilterTrans}
%\caption{Transmission of the circular dichroism filter.}
%\end{figure}

%\begin{figure}[htb]
%\centering
%\includegraphics[width=0.45 \columnwidth]{../images/WolfgrammPurity.pdf}
% \label{fig:DichroFilterTrans}
%\caption{Transmission of the circular dichroism filter.}
%\end{figure}

\subsection{Induced dichroism atomic filter}
\label{sec:InducedDichroFilter}

\btext{As with the Type-I OPO described earlier, only a small portion of the SPDC output of the Type-II OPO is atom-resonant, and a narrow-band filter is necessary to select this atom-resonant portion. } We use an induced dichroism atomic filter, similar in several ways to the FADOF of Section \ref{sec:FADOF}, to separate the frequency-degenerate output of the CESPDC from the rest of the output.  The filter, described in references \cite{CereOL2009,WolfgrammPRL2011}, has an 80 MHz FWHM passband centered on $\omNOON$ and $>$ 35 dB out-of-band rejection, so that only atom-tuned photons are detected. A representative spectrum is shown in Fig. \ref{fig:DichroFilterTrans}.

As shown in Fig.~\ref{fig:DichroSetups} (right), a YVO$_4$ crystal separates
horizontally and vertically polarized photons by 1 mm. The
polarization modes travel parallel to each other through a hot
rubidium cell of isotopically pure $^{87}$Rb, optically pumped by
a single-frequency laser resonant to the F=2$\rightarrow$F'=3
transition of the D$_2$ line of $^{87}$Rb (not shown). Due to
Doppler shifts, the optical pumping only effects a portion of the
thermal velocity distribution, and creates a circular dichroism
with a sub-Doppler linewidth of about 80~MHz. A second YVO$_4$
crystal introduces a second relative displacement, which can
re-combine or further separate the photons, depending on
polarization. Separated photons are collected, while re-combined
photons are blocked. A half wave plate is used to switch between
the ``active'' configuration, in which only photons that change
polarization in the cell are collected, and the ``inactive''
configuration, in which photons that do not change are collected.
In the ``active'' configuration, the system acts as an IFM
detector for polarized atoms: a photon is collected only if it
experiences a polarization change, i.e., if it is resonant with
the optically pumped atoms, which absorb one circular component of
the photon polarization state. Neighboring modes of the degenerate
mode at the rubidium transition are already 490~MHz detuned and
therefore outside of the filter linewidth of 80~MHz. The
out-of-band extinction ratio is $\geq$ 35~dB. The filter
transmission is optimized by adjusting the overlap between pump
and single-photon mode, the rubidium vapor temperature and the
magnitude of a small orienting applied magnetic field. The
temperature is set to 65$^{\circ}$C, which corresponds to an
atomic density of $5\times 10^{11}$ cm$^{-3}$. The measured filter
transmission of 10.0 \% for horizontal polarization and 9.5 \% for
vertical polarization is limited by pump power and in principle
can reach 25\% \cite{CereOL2009}.

To avoid contamination of the
single-photon mode by scattered pump light, the pump enters the
vapor cell at a small angle and counter-propagating to the
single-photon mode. Interference filters centered on 795~nm
further reject the 780~nm pump light with an extinction ratio of
$> 10^{5}$. The measured contribution from pump photons is below
the detectors' dark count rate. Each output is coupled into
single-mode fiber. One is detected directly on a fiber-coupled
avalanche photo diode (APD, Perkin Elmer SPCM-AQ4C). The other is
used for subsequent experiments. Photon detections are recorded by
a counting board (FAST ComTec P7888) for later analysis.

%Linear interferometry with non-entangled states can reach at best the standard quantum limit (SQL) $\delta \phi = 1/\sqrt{N}$, where $\phi$ is the interferometric phase to be measured and $N$ is the number of probe particles. When increasing $N$ is not possible, quantum enhancement offers a practical advantage. A key example is interferometric gravitational-wave detection: in current operating conditions, squeezing improves sensitivity whereas increasing photon flux produces deleterious thermal effects \cite{GEO2011}. Here we study an analogous number-limited scenario with very broad potential application: the probing of delicate systems, i.e., material systems which suffer significant damage due to the probing process. Examples are found in atomic \cite{Tey2008}, molecular \cite{Pototschnig2011}, condensed matter \cite{Eckert2008} and biological \cite{Carlton2010} science.

%\mtext{By the Kramers-Kronig relations, interferometric phase shifts are necessarily accompanied by absorption, implying deposition of energy in the probed medium.  Moreover, loss reduces or even reverses quantum sensitivity advantages\cite{Kacprowicz2010,ThomasPeterPRL2011,Escher2011}.  In real media, losses depend on both the state of the probe photons and on the state of the medium, behaviours not yet considered in the literature. To test a fully-realistic scenario, we probe a precisely understood material system with a quantum state permitting rigorous sensitivity and damage analysis.}

\subsection{Spectral purity measurement}
\label{sec:PurityII}

\begin{figure}[htb]
\centering
\includegraphics[width=0.44 \columnwidth]{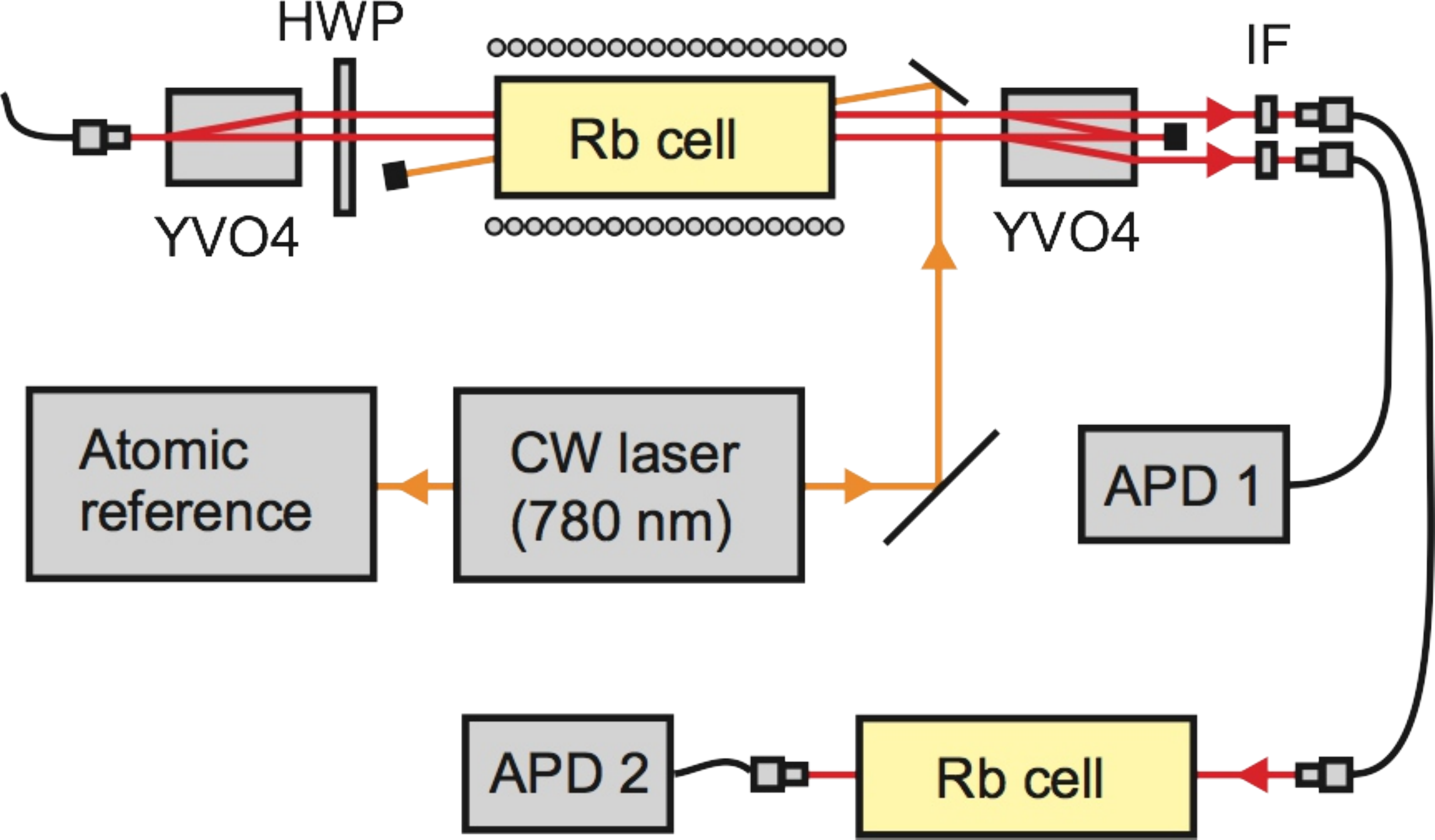}
\hspace{\fill}
\includegraphics[width=0.52 \columnwidth]{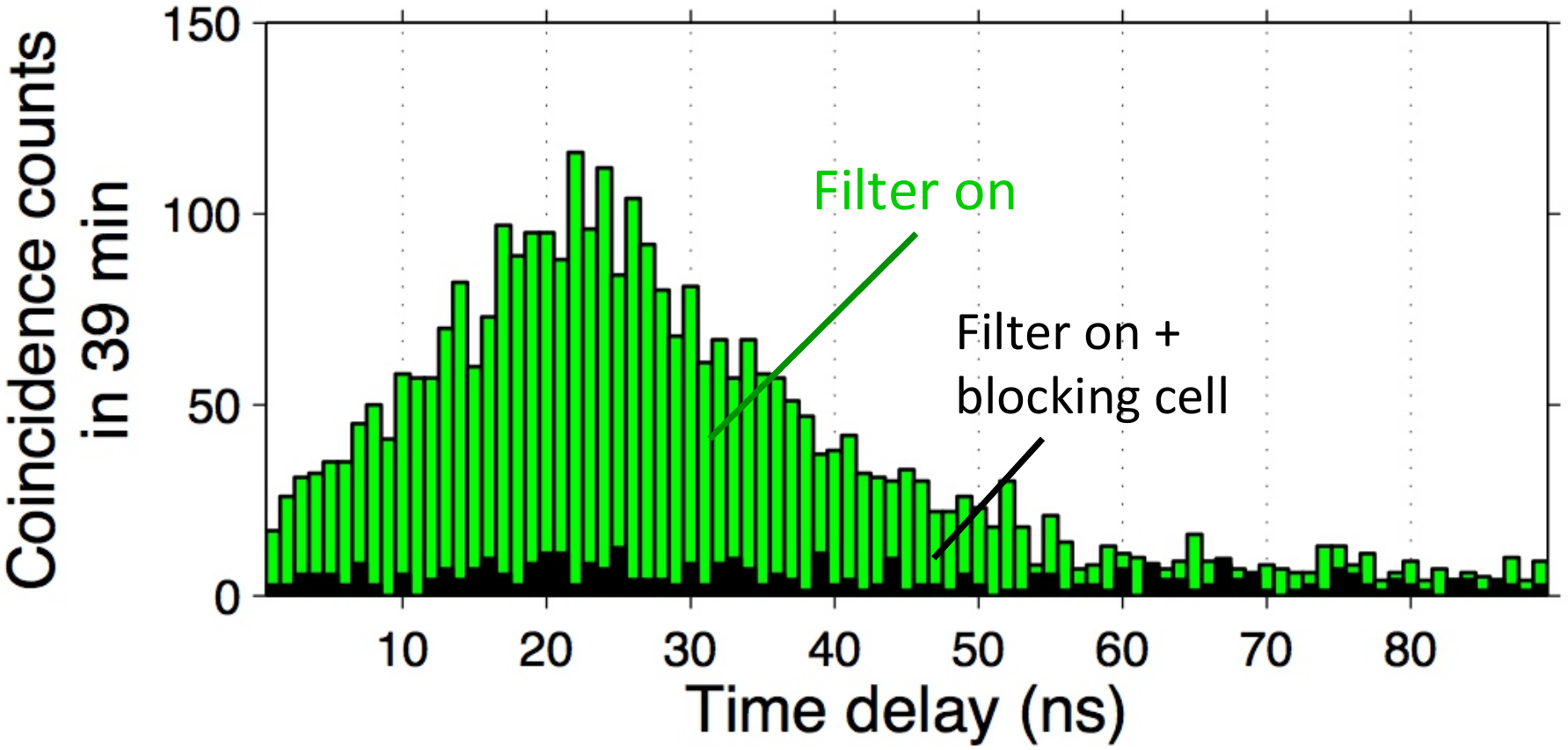}
\caption{Induced dichroism filter spectral purity measurement.  (left) schematic of the setup.  (right) arrival-time correlations with the second vapor cell cold (green) and hot (black).  When the second vapor cell is hot, it blocks the output to the dark count level of the detectors, implying at least 94\% atom-resonant photons. }
 \label{fig:DichroSpectralPurity}
\end{figure}

The CESPDC source, shown in Fig. \ref{fig:TypeIISetup}, was filtered using the geometry of Fig. \ref{fig:DichroSetups} (right), to produce in-principle spectrally pure photon pairs.  The geometry of the filter and detection setup are shown in Fig. \ref{fig:DichroSpectralPurity} (left).  
To test the spectral purity, one photon was detected as a herald, and the other subjected to further spectral filtering using a cell of Rb vapor.  The cell could be maintained at room temperature, causing little absorption, or at a high temperature, in which case it efficiently blocked resonant photons.  Fig. \ref{fig:DichroSpectralPurity} (right) shows the results: with the cold blocking cell, a double-exponental arrival-time distribution is observed, as expected.  When the blocking cell is heated, the coincidences drop to the dark-count level.  We estimate the fraction of atom-resonant photons among the heralded single photons is at least 94 \% \cite{WolfgrammPRL2011}. 

\subsection{Quantum-enhanced sensing of atoms using atom-tuned \noon states.    }
\label{sec:QESNooN}

\begin{figure}[t]
\centering
\includegraphics[width=0.95 \columnwidth]{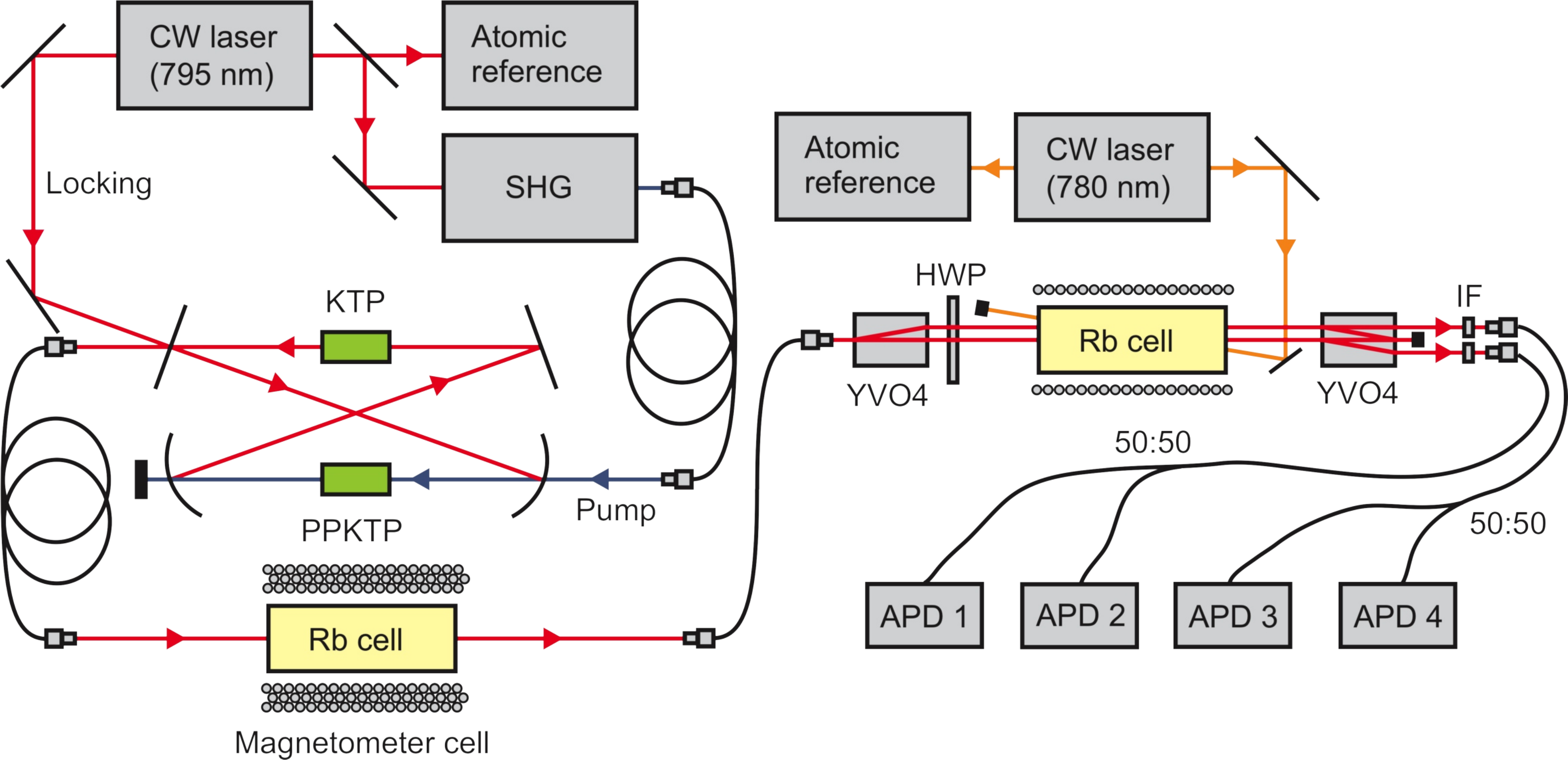}
\caption{Schematic of the setup for probing of an atomic magnetometer with atom-resonant \noon states.  Photon pairs are generated by CESPDC as in Fig. \ref{fig:TypeIISetup}, passed through a heated cell containing isotopically-enriched $^{85}$Rb and immersed in a longitudinal magnetic field, filtered as in Fig. \label{fig:DichroSetups} (right), and detected with a pair of Hanbury-Brown and Twiss setups. The three possible coincidence outcomes:  $HH$, $HV$, and $VV$ can then be monitored as a function of the applied B field.}
 \label{fig:NooNMagSetup}
\end{figure}

We now apply the atom-tuned \noon states for sensing of Faraday rotation in a hot atomic ensemble.  Because the Faraday rotation is a resonant phenomenon, it is essential to have near-resonant photons for this purpose.  We use a $^{85}$Rb atomic spin ensemble, similar to ensembles used for optical quantum memories \cite{JulsgaardN2004} and quantum-enhanced atom interferometry
\cite{SewellPRL2012, SewellNP2013, SewellPRX2014}. Non-destructive dispersive measurements on these systems, used for storage and readout of quantum information or to produce spin squeezing,
are fundamentally limited by scattering-induced depolarization \cite{MadsenPRA2004,JulsgaardN2004,KoschorreckPRL2010b,NapolitanoN2011}. This provides an experimentally-grounded motivation for the idea that the number of probe photons is a limiting resource when measuring this system; the number of probe photons cannot be increased without increasing the damage to the spin ensemble.

The setup is shown schematically in Fig. \ref{fig:NooNMagSetup}. Narrowband \noon states at $\omNOON$, the optical frequency of the $5^{2}{\rm S}_{1/2} {\rm F}$=$2 \rightarrow 5^{2}{\rm P}_{1/2} {\rm F}'$=$1$ transition of the D$_1$ line of $^{87}$Rb, are generated by CESPDC, as described above, and sent through the ensemble. The ensemble of $^{85}$Rb atoms is contained in an anti-reflection coated vapor cell with internal length $L=75$ mm, in a temperature-controlled oven at 70$^\circ$ C, together with a 0.5\% residual $^{87}$Rb component. An applied axial magnetic field $B$ of up to 60 mT produces resonantly-enhanced Faraday rotation of the optical polarization. After leaving the vapor cell, the photons are separated in polarization, frequency filtered, and
detected with single-photon counters. Two counters on each polarization output record all possible outcomes, i.e., { coincidences of} HH, HV, and VV polarizations.}

\subsubsection{Physics of near-resonant Faraday rotation }

\begin{figure}[t]
\centering
\includegraphics[width=0.8\textwidth]{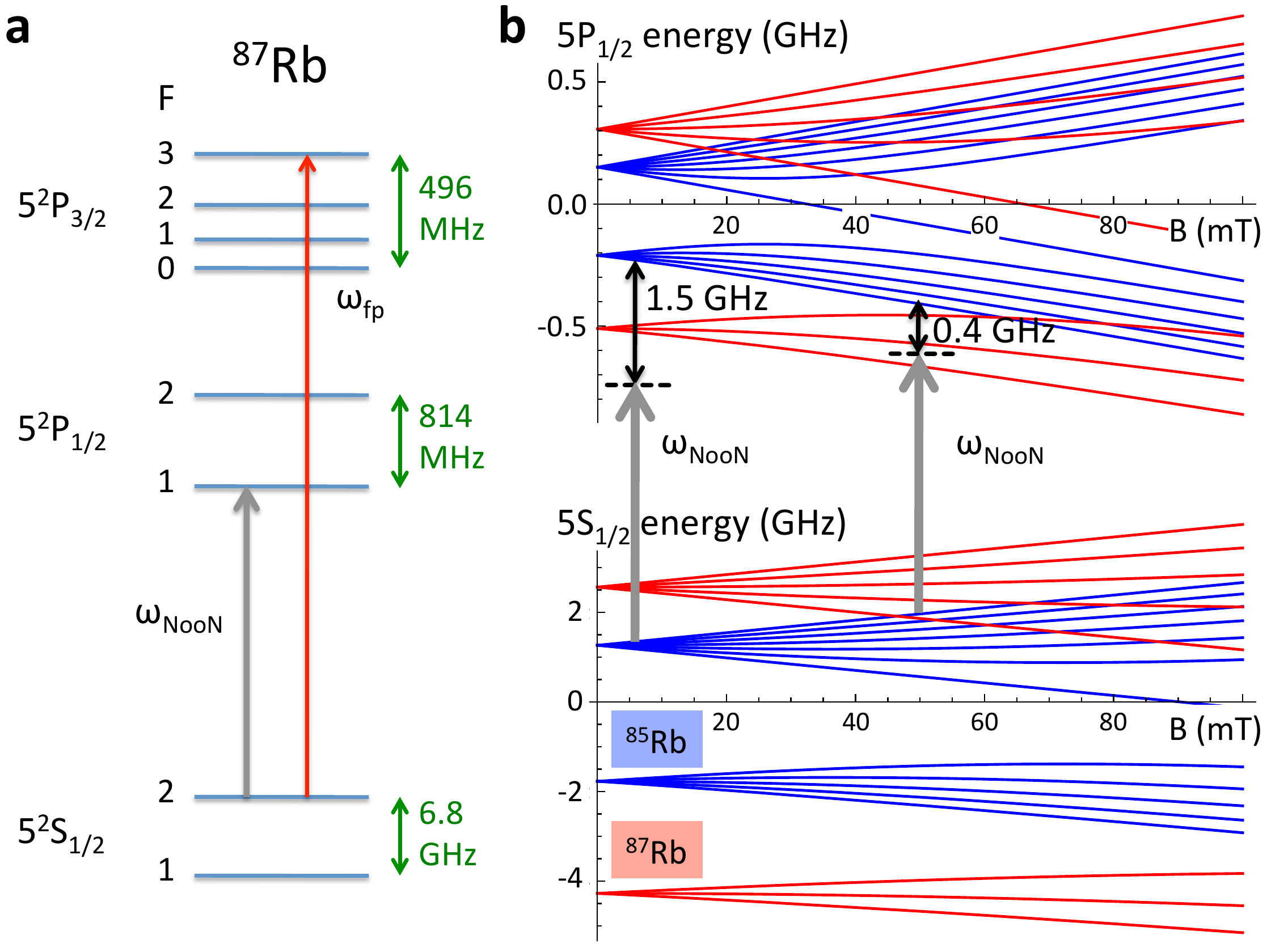}
%\hspace{\fill}
%\includegraphics[width=0.5 \columnwidth]{../images/WolfgrammMagSpectrum.pdf}
\caption{Energy level diagrams relevant to Faraday rotation on the D$_1$ line of Rb.   {\bf a} Energy levels of $^{87}$Rb relevant to generation and filtering (not to scale). The frequency of the \noon state $\omega_{\rm NooN}$ is tuned to the $5^{2}{\rm S}_{1/2} {\rm F}$=$2 \rightarrow 5^{2}{\rm P}_{1/2} {\rm F}'$=$1$ transition of the D$_1$ line of $^{87}$Rb. The optical pumping laser of the filter, with frequency $\omega_{\rm fp}$, addresses the $5^{2}{\rm S}_{1/2} {\rm F}$=$2 \rightarrow 5^{2}{\rm P}_{3/2} {\rm F}'$=$3$ transition of the D$_2$ line of $^{87}$Rb. The 15 nm separation from the detection wavelength allows a high extinction using interference filters centered on $\omega_{\rm NooN}$. {\bf b} D$_1$ energy levels of the probed ensemble versus field strength $B$, showing $^{85}$Rb levels in blue and $^{87}$Rb levels in red.  At zero field $\omega_{\rm \noon}$ is 1.5 GHz detuned from the nearest $^{85}$Rb transition.  With increasing $B$, the nearest $^{85}$Rb transition moves closer to resonance, increasing the Faraday rotation.  The Doppler-broadened absorption begins to overlap $\omega_{\rm \noon}$ near $B=$ {50} {mT}. 
\label{fig:ED}}
\end{figure}

%\begin{figure}[h!]
%\centering
%%\includegraphics[width=0.5\textwidth]{../images/Wolfgramm_fig8.pdf}
%\includegraphics[width=0.5 \columnwidth]{../images/WolfgrammMagSpectrum.pdf}
%\caption{Relevant energy level diagrams.   {\bf a} Energy levels of $^{87}$Rb relevant to generation and filtering (not to scale). The frequency of the NooN state $\omega_{\rm NooN}$ is tuned to the $5^{2}{\rm S}_{1/2} {\rm F}$=$2 \rightarrow 5^{2}{\rm P}_{1/2} {\rm F}'$=$1$ transition of the D$_1$ line of $^{87}$Rb. The optical pumping laser of the filter, with frequency $\omega_{\rm fp}$, addresses the $5^{2}{\rm S}_{1/2} {\rm F}$=$2 \rightarrow 5^{2}{\rm P}_{3/2} {\rm F}'$=$3$ transition of the D$_2$ line of $^{87}$Rb. The 15 nm separation from the detection wavelength allows a high extinction using interference filters centered on $\omega_{\rm NooN}$. {\bf b} D$_1$ energy levels of the probed ensemble versus field strength $B$, showing $^{85}$Rb levels in blue and $^{87}$Rb levels in red.  At zero field $\omega_{\rm \noon}$ is 1.5 GHz detuned from the nearest $^{85}$Rb transition.  With increasing $B$, the nearest $^{85}$Rb transition moves closer to resonance, increasing the Faraday rotation.  The Doppler-broadened absorption begins to overlap $\omega_{\rm \noon}$ near $B=$ {50} {mT}. 
%\label{fig:ED}}
%\end{figure}

\begin{figure}[t]
\centering
\includegraphics[width=0.75 \columnwidth]{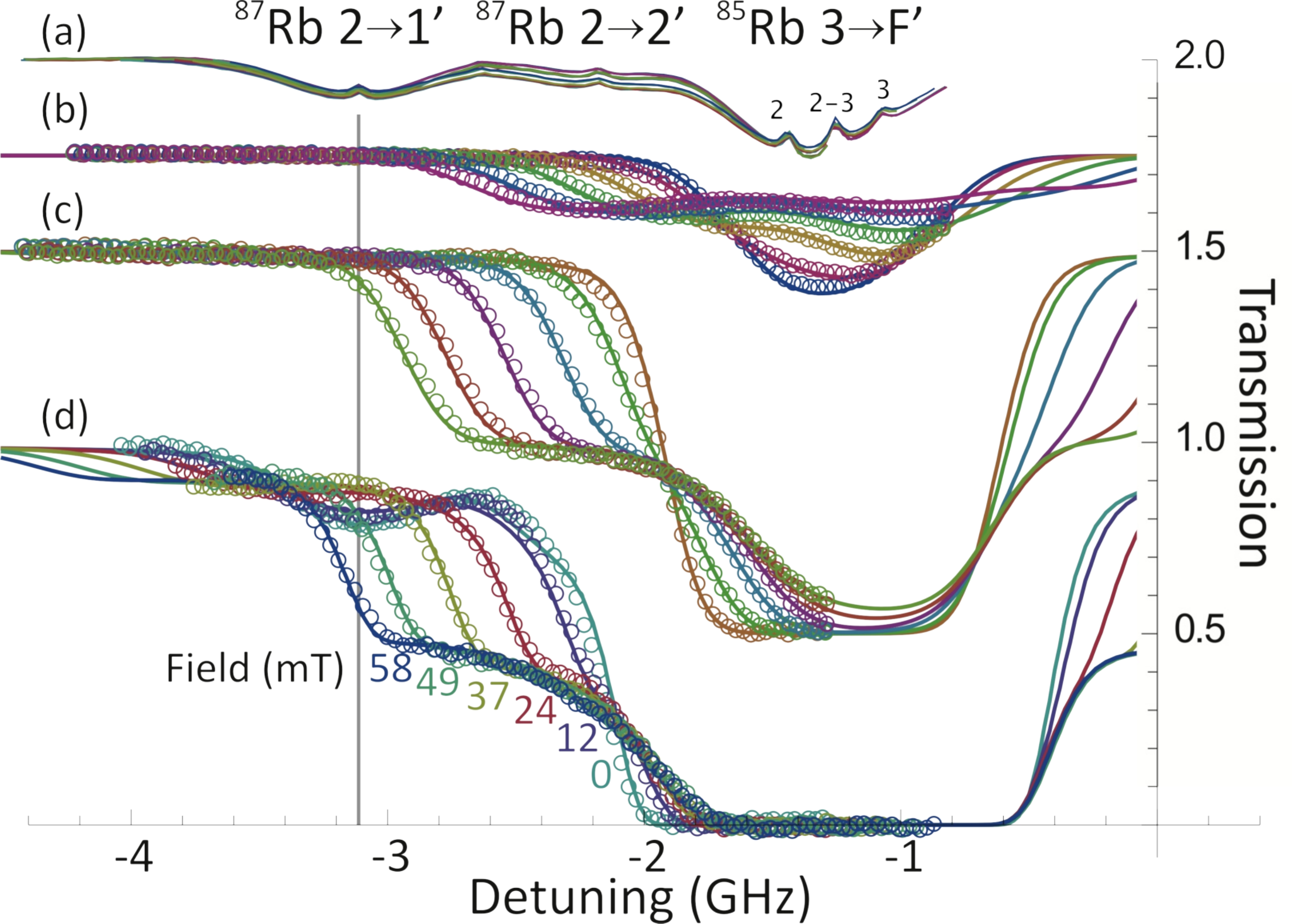}
\caption{Spectroscopic characterization of the rubidium atomic
ensemble. Circles show measured values, curves show predictions of
first-principles model (see text). {\bf a} Saturated-absorption
spectra acquired with a natural-abundance cell at room
temperature, as a frequency reference. Horizontal axis shows
detuning from the center of the $D_1$ spectral line. {\bf b}-{\bf
d} Transmission spectra for the cell containing $^{85}$Rb plus 0.5\% $^{87}$Rb at
temperatures of $22^\circ$C, $53^\circ$C and $83^\circ$C,
respectively. For each temperature, spectra with measured field
strengths (in mT) of 0, 12, 24, 37, 49, and 58 are shown, in order
of increasing line broadening. Grey vertical line shows $\omNOON$,
the probe detuning. This operating point gives strong Faraday
rotation with low absorption over the range 0-49 mT. Absorption
from the small residual $^{87}$Rb component can be seen
in {\bf d}. For clarity, parts {\bf a}-{\bf c} have been offset
vertically by 1, 0.75, and 0.5, respectively.
}
 \label{fig:DichroFilterTrans}
\end{figure}

%In contrast to quantum-enhanced probing of atoms using squeezed light \cite{Wolfgramm2010a} that did not demonstrate low-damage probing, the use of entangled photons in the single-photon regime allows for a rigorous quantification of the damage, an information-per-damage ratio close to the Heisenberg limit, the quantification of the metrological advantage by the application of Fisher information theory and hence the first low-damage probing of a delicate system.

%We probe the ensemble with a polarization \noon state, a two-mode entangled state of the form $|N\rangle_L|0\rangle_R + \exp[i \phi] |0\rangle_L|N\rangle_R$ (normalization omitted), in which $N$ particles are either all $L$- or all $R$-circularly 
%polarized. \mtext{NooN states with up to five photons have been produced \cite{Afek2010}, larger NooN states have been made from smaller ones \cite{Mitchell2004} and heralded NooN states  \cite{Smith2008,Matthews2011} have been demonstrated. To this NooN technology, we add tunable, narrow-band NooN states, suitable for efficient interaction with atoms.}
%{The use of entangled photons in the single-photon regime allows a rigorous quantification, by quantum state tomography and Fisher Information (FI) theory, of the information gained.  

%

\begin{figure}[t]
\centering
\includegraphics[width=0.7\textwidth]{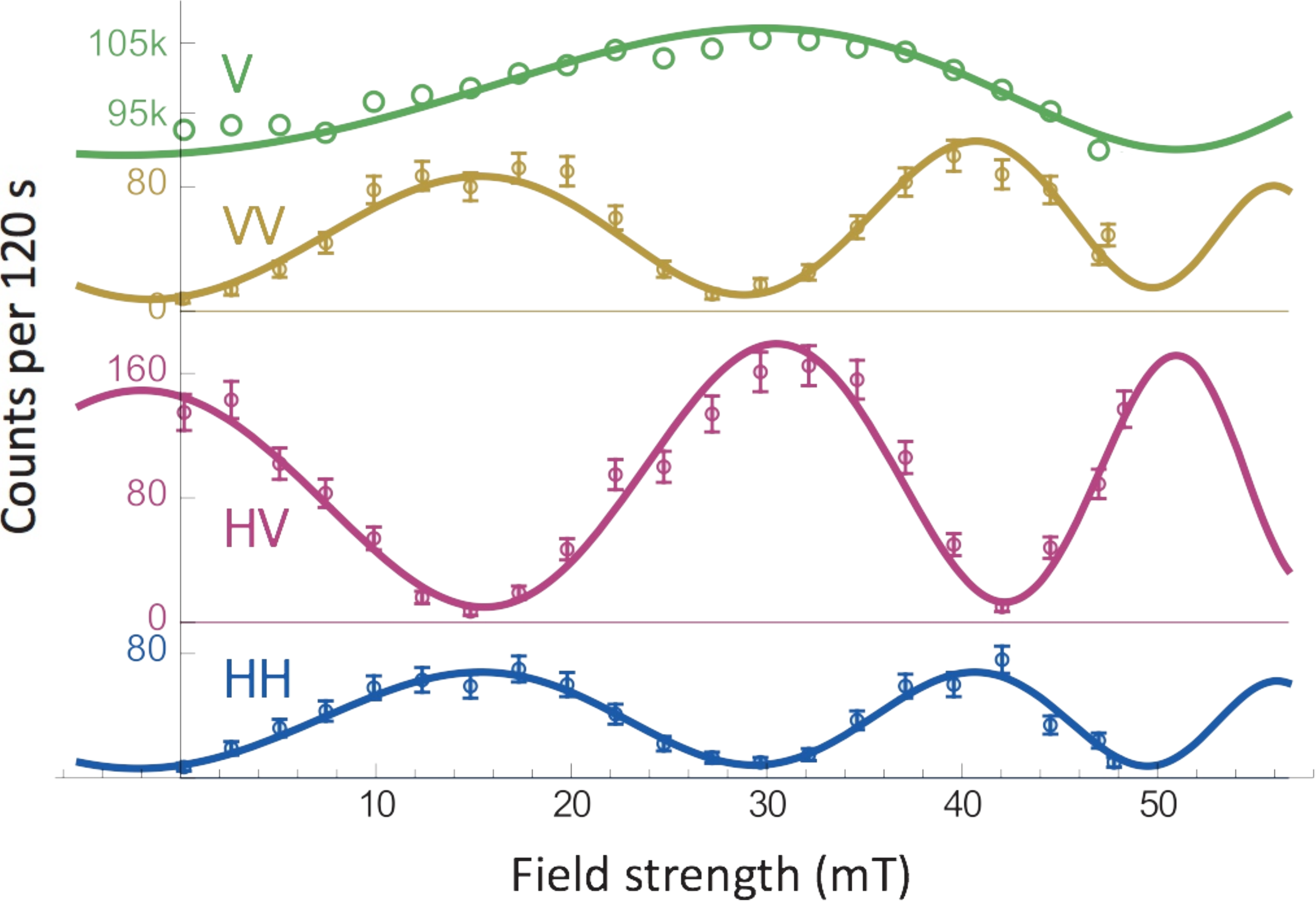}
\caption{
High-visibility super-resolving Faraday rotation probing using optical \noon states. Top curve: for phase reference, singles detection rate (V polarization) {versus field strength B} 
shows one oscillation in the range B=0 to 50 mT. Other curves: coincidence detections HH, HV, and VV show two oscillations in the same range (super-resolution) and high visibility. Symbols show measured data (no background subtracted), with $\pm 1 \sigma$ statistical uncertainties. Curves show predicted coincidence rates using a first-principles atomic susceptibility calculation illustrated in Fig. \ref{fig:ED} with a \noon state $\rhoin$ found by quantum state tomography.}
\label{fig:SuperRes}
\end{figure}

The cell, with an internal path of 75 mm and containing purified
$^{85}$Rb with a small ($0.5$\%) admixture of $^{87}$Rb, no buffer
gas, and no wall coatings that might preserve polarization, is
modeled as a thermal equilibrium, Doppler-broadened vapor subject
to Zeeman shifts in the intermediate regime. The
atomic structure is calculated by diagonalization of the atomic
Hamiltonians $H_{\rm At}\supiso = H_{0}\supiso + H_{\rm
HFS}\supiso + H_{\rm Z}\supiso$, where $H_{0}\supiso$ is the
energy structure of the isotope $^{\rm iso}$Rb including fine-structure
contribution, $H_{\rm HFS}\supiso = g_{\rm HFS} \bJ \cdot \bI$ is
the hyperfine contribution, and $H_{\rm Z}\supiso = \bB \cdot(
g_{J} \bJ + g_{I} \bI )$ is the Zeeman contribution. All atomic parameters
are taken from references 
\cite{steckRb85}, \cite{steckRb87}. 
{The matrices $H_{\rm At}\supiso$ are numerically diagonalized to find} 
field-dependent energy eigenstates, {illustrated in Fig. \ref{fig:ED},} from which the complex linear
optical polarizability is calculated, including radiative damping.  
The complex refractive index $n_\pm$ for $\sigma_\pm$
polarizations is computed including Doppler broadening and a
temperature-dependent atom density given by the vapor pressure
times the isotope fraction, and the transfer function for the cell is calculated 
from the integral of the index along the beam path, including the measured 
drop in field strength of 15\% from the center to the faces of the cell.  Transmission
spectroscopy, shown in Fig. \ref{fig:DichroFilterTrans}, agrees well with theory.

Figs. \ref{fig:ED} and  \ref{fig:DichroFilterTrans} illustrate the optical physics of this magneto-optic rotation:  The probe photons are red detuned from the $^{85}$Rb transitions, and experience the same positive contribution to the refractive index at zero field.  When the B-field is applied, however, the nearby $^{85}$Rb $F=3 \rightarrow F'$ transitions split, with the now circularly-polarized transitions moving closer or farther from the probe frequency in function of their polarization.  This provides a growing refractive index contribution for one circular polarization, and a decreasing contribution for the other, i.e. a circular birefringence giving rise to polarization rotation.  Due to the proximity to resonance, the rotation angle increases non-linearly with B until at around B = 50 mT the $^{85}$Rb $F=3 \rightarrow F'$ lines begin to overlap with the probe frequency and significant absorption begins.  

\subsubsection{Faraday rotation signals with atom-tuned \noon states}

As seen in Fig. \ref{fig:SuperRes}, all coincidence outcomes
oscillate as a function of $B$, with two-fold super-resolution relative to the single-photon oscillation, visible in the singles counts due to a small imbalance between H and V in the input state. The interference visibilities are all $\ge90$\%, well above the 33\% classical limit for HH and VV visibility \cite{AfekPRL2010}. Also shown are predicted coincidence rates \cite{WolfgrammNPhot2013}, which show good agreement.  

An analysis of the Fisher information from these coincidence rates confirm the utility of entangled states for probing atomic ensembles.  The \noon state achieves a factor 1.30$\pm$0.05 more Fisher information per photon than the standard quantum limit (SQL).  I.e. it gives more information per photon than can be extracted using any state consisting of non-entangled photons.  It achieves this advantage at a field of $B=37$ mT, near the point at which absorption begins to become important.  Perhaps more important to eventual application, the  \noon state gives also an advantage in Fisher information per scattered photon, the figure of merit for low-damage probing. 
As described in \cite{WolfgrammNPhot2013} the \noon state gives an advantage of $1.40 \pm$ 0.06 over the SQL.

%We quantify scattering from the $^{85}$Rb ensemble as $\scat = {\rm Tr} [\rho \Pi_{\rm scat}]$, where $\Pi_{\rm scat} \equiv {\rm diag}(\lpp,\lpm,\lmp,\lmm)$ in the $\sigma_{\pm}$ basis and the mean number of scattering events is $s_{ab} \equiv 2-|t_a^{(85)}|^2-|t_b^{(85)}|^2$, where $t_\pm^{(85)}$ is the $^{85}$Rb contribution to $t_\pm$. A completely analogous calculation is made for single-photon scattering. As above, the the SQL for this figure of merit is found by numerical optimization. 

Among other things, this exercise uncovered a previously unknown feature of quantum-enhanced sensing in imperfect interferometers, in this case those with losses.  Here, as one can expect for most interferometric measurements on material systems, the loss depends on the measured quantity (here $B$). This dependence makes a positive contribution to the Fisher information, offsetting the well-known \cite{ThomasPeterPRL2011} reduction of Fisher information due to loss of photons.  

\section{Conclusions}

We have described two experimental projects to generate high-spectral purity indistinguishable photon pairs using cavity-enhanced SPDC and extremely narrowband optical filters based on resonant optical effects in atomic ensembles.  Combining these techniques, we have demonstrated bright sources of entangled photons pairs, with both high two-photon coherence, for example 99\% fidelity with a two-photon \noon state, and high spectral purity, $\ge$ 94\% atom-resonant heralded single photons for a type-II source and $\ge$ 98\% atom resonant photon pairs for a type-I source.  The potential for interaction with atoms is clearly shown by the generation and use of atom-tuned \noon states to beat the standard quantum limit in non-destructive probing of an atomic magnetometer.  
As this work goes forward, it will be interesting to see how  indistinguishable photon pairs can interact with other atomic systems, for example atomic quantum memories and atomic quantum information processors.

\section*{{Appendix: Second-order correlation functions of filtered output}}
\label{sec:FilteredCorrFunctions}

In this section we consider the second order correlation function of the field operators $\aout$ in a form:
\begin{equation}
   G^{(2)}(T) \propto \langle \aout^\dagger(t) \aout^\dagger(t+T) \aout(t+T) \aout(t) \rangle 
\end{equation}
for multimode (unfiltered) and single-mode (filtered) output of the OPO.

As shown  by Lu et al. \cite{LuPRA2000}, $G^{(2)}(T)$ describing the output of a single-mode, far-below-threshold OPO has the form of double exponential decay
\begin{equation}
\label{g2single}
   G_{\rm single}^{(2)}(T)\propto e^{-|T|(\gamma_1+\gamma_2)},
\end{equation}
where the reflectivity of the output coupler is $r_1=\exp[{-\gamma_1 \tau}]$, the effective reflectivity resulting from intracavity losses is $r_2=\exp[{-\gamma_2 \tau}]$ and $\tau$ is the cavity round-trip time.  
An ideal narrowband filter would remove all the nondegenerate cavity-enhanced spontaneous down-conversion {CESPDC} 
%{OPO}
 modes, reducing the $G^{(2)}(T)$  to $G_{\rm single}^{(2)}(T)$.  This filtering effect was demonstrated in \cite{WolfgrammPRL2011} for a type-II OPO and an induced dichroism atomic filter.

In \cite{LuPRA2000} it is also predicted that when the filter is off, so that the output consists of $N$ cavity modes,  $G_{}^{(2)}(T)$ takes the form
\begin{eqnarray}
\label{eq:GmultiSin}
   G_{\rm multi}^{(2)}(T) &\propto & G_{\rm single}^{(2)}(T) \frac{\sin^2[(2N+1)\pi T/ \tau]}{(2N+1)\sin^2[\pi T/\tau]}
 \\ 
& \approx &  G_{\rm single}^{(2)}(T)\sum_{n=-\infty}^{\infty}\delta(T-n\tau),
\end{eqnarray}
i.e., with the same double exponential decay but modulated by a comb with a period equal to the cavity round-trip time $\tau$.  In our case the bandwidth of the output contains more than 200 cavity modes, and the fraction in Eq. (\ref{eq:GmultiSin}) is well approximated by a comb of Dirac delta functions. 

The comb period of $\tau = 1.99$~ns is comparable to the $t_{\rm bin} = 1$ ns resolution of our counting electronics, a digital time-of-flight counter (Fast ComTec P7888).  This counter assigns arrival times to the signal and idler arrivals relative to an internal clock.  We take the ``window function'' for the $i$th bin, i.e., the probability of an arrival at time $T$ being assigned to that bin,  to be
\begin{equation}
   f^{(i)}(T)=
\begin{cases}
    1,& \text{if } T\in [i t_{bin}, (i+1) t_{bin}] \,,   \\
    0,              & \text{otherwise}\,.
\end{cases}
\end{equation}
 
Without loss of generality we assign the signal photon's bin as $i=0$, and we include an unknown relative delay $T_0$ between signal and idler due to path length, electronics, cabling, and so forth.  For a given signal arrival time $t_s$, the rate of idler arrivals in the $i$th bin is $\int dt_i \,  f^{(i)}(t_i) G_{\rm multi}^{(2)}(t_i - t_s - T_0)$ ($t_i$ is the idler arrival time).  This expression must be averaged over the possible $t_s$ within bin $i=0$.  We also include the ``accidental'' coincidence rate $G_{\rm acc}^{(2)}= t_{\rm bin} R_1 R_2$, where $R_1, R_2$ are the singles detection rates at detectors $1,2$, respectively. The rate at which coincidence events are registered with $i$ bins of separation is  then 
\begin{eqnarray}
   G_{\rm multi,det}^{(2)}(i) &= & \frac{1}{t_{\rm bin}} \int  dt_s \, f^{(0)}(t_s) \int dt_i \,  f^{(i)}(t_i) G_{\rm multi}^{(2)}(t_i - t_s - T_0) +  G_{\rm acc}^{(2)} \\
   &=&  \sum_{n=-\infty}^{\infty} G_{\rm single}^{(2)}(n \tau) \frac{1}{t_{\rm bin}} \int_0^{t_{\rm bin}} dt_s \,  f^{(i)}(t_s + T_0 + n \tau)  +  G_{\rm acc}^{(2)}. 
\end{eqnarray}

 We take $T_0$ is a free parameter in fitting to the data. Note that if we write $T_0=k t_{\rm bin}+\delta$ then the simultaneous events fall into $k$th bin and $\delta\in[-t_{\rm bin}/2,t_{\rm bin}/2]$ determines where the histogram has the maximum visibility due to the beating between the 1 ns sampling frequency of the detection system and the 1.99 ns comb period.   APD time resolution is estimated to be $350$ ps FWHM (manufacturer's specification), i.e. significantly less than the TOF uncertainty, and is not included here.

\addcontentsline{toc}{section}{Appendix}

\begin{acknowledgement}
The work reported here involved many people over many years.  I would especially like to thank Marta Abad, Federica Beduini, Alessandro Cer\`{e}, Nicolas Godbout, Valentina Parigi, Ana Predojevi\'{c}, Chiara Vitelli, Florian Wolfgramm, Xinxing Xing, and Joanna Zieli\'{n}ska, each of whom contributed something unique and essential to the work reported here, e.g. inventing a first-of-its-kind source or filter, providing insights into the physics of atomic optical instruments, or persuading difficult lasers (and their suppliers) to cooperate with our plans.  Aephraim Steinberg was essential to getting the photon pair research started.  The contributions of Zehui Zhai, Yannick de Icaza Astiz and Gianvito Lucivero are also much appreciated.  The research was supported by various Catalan, Spanish, European, Canadian and philanthropic grants over the years.  The writing of this chapter was supported by the Spanish MINECO  project MAGO (Ref. FIS2011-23520), by the European Research Council project AQUMET, and by Fundaci\'{o} Privada CELLEX. 

\end{acknowledgement}

\bibliographystyle{SpringerPhysMWM} %Modification of nature.bst by MWM
\bibliography{../biblio/BigBib130123.bib,../biblio/FADOF4,../biblio/01_Quantum_OpticsBD2.03,../biblio/FDC_bib,../biblio/2ph_wavefunction.bib
%../../../Dropbox/CommonLatexFiles/BigBib130123,
}

\printindex
\end{document}